# Finite-Dimensional Bounds on $\mathbb{Z}_m$ and Binary LDPC Codes with Belief Propagation Decoders

Chih-Chun Wang, Sanjeev R. Kulkarni, H. Vincent Poor

*Abstract*— This paper focuses on finite-dimensional upper and lower bounds on decodable thresholds of $\mathbb{Z}_m$ and binary low-density parity-check (LDPC) codes, assuming belief propagation decoding on memoryless channels. A concrete framework is presented, admitting systematic searches for new bounds. Two noise measures are considered: the Bhattacharyya noise parameter and the soft bit value for a maximum *a posteriori* probability (MAP) decoder on the uncoded channel. For $\mathbb{Z}_m$ LDPC codes, an iterative $m$-dimensional bound is derived for $m$-ary-input/symmetric-output channels, which gives a sufficient stability condition for $\mathbb{Z}_m$ LDPC codes and is complemented by a matched necessary stability condition introduced herein. Applications to coded modulation and to codes with non-equiprobable distributed codewords are also discussed.

For binary codes, two new lower bounds are provided for symmetric channels, including a two-dimensional iterative bound and a one-dimensional non-iterative bound, the latter of which is the best known bound that is tight for binary symmetric channels (BSCs), and is a strict improvement over the bound derived by the channel degradation argument. By adopting the reverse channel perspective, upper and lower bounds on the decodable Bhattacharyya noise parameter are derived for non-symmetric channels, which coincides with the existing bound for symmetric channels.

## I. INTRODUCTION

The belief propagation (BP)/sum-product algorithm [1] is one of the major components in modern capacity-approaching codes, including turbo codes [2], low-density parity-check (LDPC) codes [3], repeat accumulate (RA) codes [4], etc. The BP algorithm uses distributed local computation to approximate the global maximum likelihood in an efficient way [5]. The density evolution (DE) method is a tool for explicitly computing the asymptotic behavior under iterative decoding with the assumption of independently distributed messages, which can be justified by the cycle-free convergence theorem in [6]. In each iteration, the DE method focuses on the density of the log likelihood ratio[1] (LLR), which is of infinite dimension and is a sufficient statistic completely describing arbitrary binary-input memoryless channels.

Even after the efficient implementation of density evolution by moving into the LLR domain, a one-dimensional iterative formula (or at most finite-dimensional formulae) to approximate the density evolution is very appealing since it reduces significantly the computational complexity of code degree optimization [7]. Several approximation formulae have been proposed including Gaussian approximations [8], [9], binary erasure channel (BEC) approximations, reciprocal channel approximations [10], and the EXtrinsic Information Transfer (EXIT) chart analysis [11]. The finite dimensionality also helps in the analysis of the behavior of the message passing decoder [12], [13].

Contrary to the approximations, rigorous iterative upper and lower bounds generally sacrifice the threshold predictability for specific channel models in exchange for guaranteed universal performance for arbitrary channel models. Many results have been found for binary-input/symmetric-output (BI-SO) channels, including Burshtein *et al.* [13] on the soft bit value for the maximum *a posteriori* probability (MAP) decoder, Khandekar *et al.* [14] on the Bhattacharyya noise parameter, and Land *et al.* [15] and Sutskover *et al.* [16] on the mutual information. For binary-input/non-symmetric-output (BI-NSO) channels, a loose one-dimensional iterative upper bound on the Bhattacharyya noise parameter is provided in [17], which was used to derive the stability condition of BI-NSO channels and to upper bound the asymptotic convergence rate of the bit error probability. Bennatan *et al.* [18] used an iterative upper bound to derive the stability conditions for $GF(q)$-based LDPC codes when $q$ is a power of a prime number.

This paper is organized as follows. The necessary definitions and background knowledge will be provided in Section II, including the definitions of the symmetric channels, the noise measures of interest, and the LDPC code ensemble. Section III will provide the framework for the iterative bounding problem and review some existing results. A Bhattacharyya-noise-parameter bound and a pair of stability conditions will be provided for $\mathbb{Z}_m$ LDPC codes in Section IV. For binary LDPC codes, Sections V and VI are devoted to the iterative and non-iterative bounds respectively, the former of which include a one-dimensional bound for BI-NSO channels and a two-dimensional bound for BI-SO channels, while the latter of which provides the best (tightest) known bound for binary symmetric channels (BSCs). The existing bound based on the channel degradation argument [6] is also tight for BSCs, but is very loose for other channels, compared to which our bound

---

This work was supported by the Army Research Laboratory under Contract No. DAAD19-01-2-0011.

C.-C. Wang is with the School of Electrical and Computer Engineering, Purdue University, West Lafayette, IN 47906. Email: chihw@purdue.edu

S. R. Kulkarni and H. V. Poor are with the Department of Electrical Engineering, Princeton University, Princeton, NJ 08544. Email: {kulkarni, poor}@princeton.edu

[1]In the most general setting of the DE method, the quantity of interest during the iterations can be the density of any measure of the message passing decoder. Nevertheless, the density of the LLR is capable of capturing the entire behavior of the iterative decoder, since the density of the LLR is a sufficient statistic of the corresponding detection problem.

is a strict improvement and generates much tighter results for other channel models. Performance comparisons are provided in Section VII. Section VIII concludes the paper.

## II. FORMULATION

In this paper, we consider only memoryless channels with discrete input alphabets.

### A. Symmetric Channels

*1) Definition:* A BI-SO $\mathbf{X} \mapsto \mathbf{Y}$ channel is conventionally defined as a channel with binary[2] input set $\mathbf{X} = \{0, 1\}$ and real output set $\mathbf{Y} = \mathbb{R}$, such that $P(Y = y|X = 0) = P(Y = -y|X = 1)$ where $X$ and $Y$ denote the (random) channel input and output, respectively. In the literature of LDPC codes (eg., [6]), an equivalent commonly-used definition is that the BI-SO channel satisfies $dP(m) = e^m dP(-m)$, where $dP(m)$ is the density of the LLR messages, $m := \log \frac{P(Y|X=0)}{P(Y|X=1)}$, given $X = 0$.

Let $\mathbb{Z}_m := \{0, 1, \cdots, m-1\}$ denote the integer ring modulo $m$. A more general definition for $m$-ary-input/symmetric-output (MI-SO) channels is given as follows.

*Definition 1 (MI-SO Channels):* For any function $\mathcal{T} : \mathbf{Y} \mapsto \mathbf{Y}$, let
$$\mathcal{T}^k(y) = \underbrace{\mathcal{T} \circ \mathcal{T} \circ \cdots \circ \mathcal{T}}_{k}(y)$$
denote the corresponding $k$-times self-composition of $\mathcal{T}$. An $m$-ary-input channel $\mathbb{Z}_m \mapsto \mathbf{Y}$ is (circularly) symmetric if there exists a bijective transform $\mathcal{T} : \mathbf{Y} \mapsto \mathbf{Y}$ such that $\mathcal{T}^m(y) = y, \forall y \in \mathbf{Y}$ and
$$\forall x \in \mathbb{Z}_m, F(dy|0) = F(\mathcal{T}^x(dy)|x),$$
where $F(dy|x)$ is the conditional distribution of $Y \in dy$ given $X = x$. When $m = 2$, this definition collapses to that of the conventional BI-SO channel.

*Note:* There is no constraint on $\mathbf{Y}$, the range of the channel output. For example, in phase-shift keying (PSK) or quadrature amplitude modulation (QAM) scenarios, $\mathbf{Y} = \mathbb{R}^2$.

This definition of "symmetric channels" coincides with the definition of the "matched signal set" in [19]. It is worth noting that belief propagation on LDPC codes is (circularly) symmetric, since the locally optimal BP decoder behaves identically under different transmitted codewords when the same circularly shifted error pattern is received. To be more explicit, assume that a non-zero codeword $\mathbf{x} = (x_1, \cdots, x_n)$ is transmitted and the received likelihood values are $\mathbf{p}(\mathbf{y}) = (\mathbf{p}(y_1), \cdots, \mathbf{p}(y_n))$, where each coordinate $\mathbf{p}(y_i) = (P(Y_i = y_i|X_i = 0), P(Y_i = y_i|X_i = 1), \cdots, P(Y_i = y_i|X_i = m-1))$ is a vector containing the likelihood values of $x_i$ after receiving $y_i$. The circular symmetry of the belief propagation decoder is characterized by the fact that the decoding behavior under the codeword and likelihood pair $(\mathbf{x}, \mathbf{p}(\mathbf{y}))$ is identical to the case in which the all-zero codeword $\mathbf{0}$ is transmitted and the shifted likelihood $\mathbf{p}^{\overleftarrow{\mathbf{x}}}(\mathbf{y}) = (\mathbf{p}^{\overleftarrow{x_1}}(y_1), \cdots, \mathbf{p}^{\overleftarrow{x_n}}(y_n))$ is received, where

$$\mathbf{p}^{\overleftarrow{x_i}}(y_i) = (P(Y_i = y_i|X_i = x_i), P(Y_i = y_i|X_i = x_i + 1),$$
$$\cdots, P(Y_i = y_i|X_i = m-1), P(Y_i = y_i|X_i = 0),$$
$$\cdots, P(Y_i = y_i|X_i = x_i - 1)).$$

[2]Another common setting is to consider $\mathbf{X} = \{+1, -1\}$, which reflects coherent binary phase shift keying (BPSK) modulation. However, to be compatible with the algebra on which the parity check equations are defined, we assume $\mathbf{X} = \{0, 1\}$ instead of $\{+1, -1\}$.

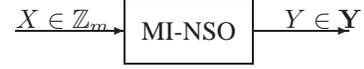

(a) An MI-NSO channel.

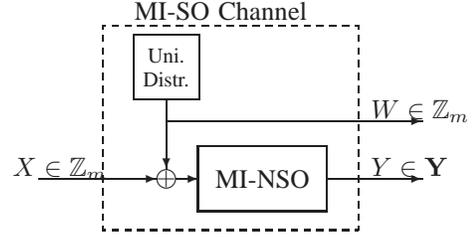

(b) A symmetrized MI-NSO channel.

Fig. 1. Channel symmetrization.

One immediate benefit of considering a symmetric channel LDPC codes is that all codewords have the same error probability under the BP decoder and we can use the all-zero codeword as a representative, which facilitates the simulation of codes with finite length. Further discussion on the BP decoder for $\mathbb{Z}_m$ LDPC codes and on the representative all-zero codeword can be found in [20] and in [19], [6], [17].

One advantage of *Definition 1* is that we can immediately prove the channel symmetrizing argument as follows. Consider an $m$-ary-input/non-symmetric-output (MI-NSO) channel in Fig. 1(a) and a concatenated new channel in Fig. 1(b), sharing the same MI-NSO channel block. Since the receiver of the latter channel is able to use the received value of $W$ to invert this concatenation, these two channels are equivalent from the detection point of view, which in turn implies that all reasonable noise measures of the two are identical, including but not limited to the channel capacities, the error probabilitiess and soft-bit values under the MAP decoder, and the Bhattacharyya noise parameters. The circular symmetry of this new equivalent channel $X \mapsto (W, Y)$ in Fig. 1(b) can then be verified by letting the bijective transform $\mathcal{T} : (\mathbb{Z}_m, \mathbf{Y}) \mapsto (\mathbb{Z}_m, \mathbf{Y})$ in *Definition 1* be $\mathcal{T}(w, y) = (w - 1, y)$. From the above discussion, Fig. 1(b) is an equivalent, symmetrized version of the original MI-NSO channel, and we can assume all channels are symmetric as long as the additional complexity of the channel symmetrizing[3] is bearable.

*2) MSC Decomposition:* One of the simplest MI-SO channels is the $m$-ary symmetric channel (MSC), which is a $\mathbb{Z}_m \mapsto \mathbb{Z}_m$ channel and can be fully specified by a parameter vector $\mathbf{p} = (p_0, p_1, \cdots, p_{m-1})$ such that the conditional probability $\mathsf{P}(Y = x+i|X = x) = p_i$, $\forall x, i \in \mathbb{Z}_m$. From *Definition 1*, it can be proved that any MI-SO channel can be uniquely expressed as a probabilistic combination of different MSCs, while $\mathbf{p}$ is observed by the receiver as side information. This decomposition is illustrated in Figs. 2(a) and 2(b), in which the probabilistic weight of different vectors $\mathbf{p}$ is denoted as $dP(\mathbf{p})$, and a formal proof of this MSC decomposition is given in APPENDIX I. When $m = 2$, an MSC collapses to a BSC and the channel-specifying vector $\mathbf{p}$ equals $(1-p, p)$, where $p$ is the crossover probability. For simplicity, we sometimes use the scalar parameter $p$ rather than the vector $\mathbf{p}$ to specify a BSC.

[3]This channel symmetrizing technique is equivalent to considering the LDPC coset code ensemble [21].

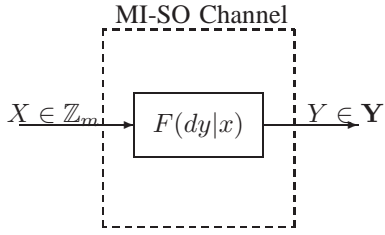

(a) By Conditional Distributions.

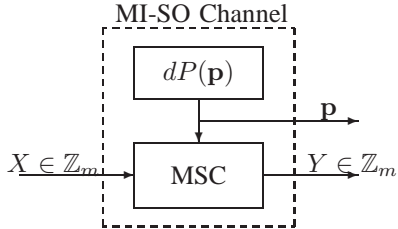

(b) By Probabilistic Combinations.

Fig. 2. Different representations for the $m$-ary-input/symmetric output channels.

*Note:* The probabilistic weight $dP(\mathbf{p})$ does not depend on the *a priori* input distribution on $\mathbf{X}$, but only depends on the channel model $\mathsf{P}(Y|X)$. This observation will be used in the proof of the non-iterative bound in Section VI.

### B. Noise Measures

*1) Binary Input Channels:* For a binary channel $\mathbf{X} = \{0, 1\} \mapsto \mathbf{Y}$, we use $p(x|y)$ to denote the *a posteriori* probability $\mathsf{P}(X = x|Y = y)$, and consider the following two noise measures:

- [The Bhattacharyya Noise Parameter (the Chernoff bound value)]

$$CB := \mathsf{E}_{X,Y} \left\{ \sqrt{\frac{p(\bar{X}|Y)}{p(X|Y)}} \right\} \quad (1)$$

$$= \sum_{x \in \{0,1\}} \mathsf{P}(X = x) \mathsf{E} \left\{ \sqrt{\frac{p(\bar{x}|Y)}{p(x|Y)}} \middle| X = x \right\},$$

where $\bar{x}$ denotes the complement of the binary input $x$. A discussion of the $CB$ parameter in turbo-like codes can be found in [22]. With uniformly distribution on $X$, $CB$ can be related to the cutoff rate $R_0$ by $R_0 = 1 - \log_2(CB + 1)$.

- [The Soft Bit Value]

$$SB := 2\mathsf{E}_{X,Y} \left\{ p(\bar{X}|Y) \right\}$$
$$= 2 \sum_{x \in \{0,1\}} \mathsf{P}(X = x) \mathsf{E} \left\{ p(\bar{x}|Y) | X = x \right\}, \quad (2)$$

which was used in the bounds of [13].

Each of the above noise measures has the property that the condition $CB = 0$ (or $SB = 0$), represents the noise-free channel, while $CB = 1$ (or $SB = 1$), implies the noisiest channel in which the output is independent of the input. It is worth noting that both $CB$ and $SB$ are well defined even for BI-NSO channels with non-uniform input distributions. Most of our theorems are derived based on the assumption of uniformly distributed $X$, and special notes will be given when non-uniform *a priori* distributions are considered.

For BSCs with uniformly distributed $X$, $CB = 2\sqrt{p(1-p)}$ and $SB = 4p(1-p)$ where $p$ is the crossover probability. By the BSC decomposition argument in Section II-A.2, the value of $CB$ or $SB$ for any BI-SO channel is simply the probabilistic average of the corresponding values of the constituent BSCs, that is, for uniformly distributed $X$

$$CB = \int 2\sqrt{p(1-p)}\, dP(p)$$
$$SB = \int 4p(1-p)\, dP(p).$$

The above formulae will be extensively used in our derivation of finite dimensional bounds. In the context of density evolution [6], $CB$ and $SB$ with uniformly distributed $X$ can be expressed as

$$CB = \int_{m=-\infty}^{\infty} e^{-\frac{m}{2}}\, dP(m)$$
$$SB = \int_{m=-\infty}^{\infty} \frac{2}{1+e^m}\, dP(m), \quad (3)$$

where $m := \log \frac{\mathsf{P}(\mathbf{Y}|X=0)}{\mathsf{P}(\mathbf{Y}|X=1)}$ is the passed LLR message and $dP(m) := d\mathsf{P}(m|X = 0)$ is the density of $m$ given $X = 0$. With the assumption of uniformly distributed $X$, the $CB$ and $SB$ values of some common channel models are given as follows (listed in order from the most BSC-like to the most BEC-like[4]):

1) The BSC with crossover probability $p$:

$$CB = 2\sqrt{p(1-p)} \triangleq CB(p)$$
$$SB = 4p(1-p) \triangleq SB(p).$$

---

[4]The order is obtained by plotting the $(CB, SB)$ values of different channel models of the same capacity in a two-dimensional plane, similar to Fig. 7. The channel models are then sorted according to their distances related to the points corresponding to a BSC and a BEC.

2) The binary-input Laplace channel (BiLC) with variance $2\lambda^2$, i.e. $p_L(y) = \frac{1}{2\lambda}\exp\left(-\frac{|y|}{\lambda}\right)$:

$$CB = \frac{1+\lambda}{\lambda}\exp\left(-\frac{1}{\lambda}\right)$$

$$SB = \frac{\exp\left(-\frac{1}{\lambda}\right)}{\cosh\left(\frac{1}{\lambda}\right)} + 2\exp\left(-\frac{1}{\lambda}\right)\arctan\left(\tanh\left(\frac{1}{2\lambda}\right)\right).$$

3) The binary-input additive white Gaussian channel (Bi-AWGNC) with noise variance $\sigma^2$:

$$CB = \exp\left(-\frac{1}{2\sigma^2}\right)$$

$$SB = \frac{1}{\sqrt{2\pi\sigma^2}}\exp\left(-\frac{1}{2\sigma^2}\right)\int_{-\infty}^{\infty}\frac{\exp\left(-\frac{x^2}{2\sigma^2}\right)}{\cosh\left(\frac{x}{\sigma^2}\right)}dx.$$

4) The binary-input Rayleigh fading channel with unit input energy and noise variance $\sigma^2$, i.e. the density function of the output amplitude is $p_A(a) = 2a\exp(-a^2)$ and the additve noise distribution is $\mathcal{N}(0,\sigma^2)$:

$$CB = \frac{1}{1+\frac{1}{2\sigma^2}}$$

$$SB = \frac{2}{\sqrt{2\pi\sigma^2}}$$
$$\cdot \int_0^{\infty}\int_{-\infty}^{\infty} a\exp(-a^2)\frac{\exp\left(-\frac{x^2+a^2}{2\sigma^2}\right)}{\cosh\left(\frac{xa}{\sigma^2}\right)}dxda.$$

5) The BEC with erasure probability $\epsilon$:

$$CB = SB = \epsilon.$$

*2) $m$-ary Input Channels:* For $m$-ary-input channels, we define the pairwise Bhattacharyya noise parameter from $x$ to $x'$ as follows:

$$CB(x \to x') := \mathsf{E}\left\{\left.\sqrt{\frac{\mathsf{P}(x'|Y)}{\mathsf{P}(x|Y)}}\right| X = x\right\}. \quad (4)$$

Considering any MI-SO channel with uniformly distributed input $X$, we immediately have

[Symmetry:] $\quad CB(x \to x') = CB(x' \to x) \quad (5)$

[Stationarity:] $\quad CB(x \to x') = CB(0 \to x' - x).$

By stationarity, we can then use $\mathsf{CB} := \{CB(0 \to x')\}_{x' \in \mathbb{Z}_m}$ as the representing vector for all $CB(x \to x')$. Also assuming the uniform distribution on $X$, the cutoff rate $R_0$ and $\mathsf{CB}$ can be related as follows [23]:

$$R_0 = \log_2 m - \log_2\left(\sum_{x' \in \mathbb{Z}_m} CB(0 \to x')\right). \quad (6)$$

Example:
- For an MSC with parameter **p** and uniformly distributed $X$, we have

$$CB(0 \to x) = \sum_{y \in \mathbb{Z}_m}\sqrt{p_y p_{y+x}}.$$

When $m = 2$, the representing vector becomes $\mathsf{CB} = (1, CB(p))$, where $CB(p) = 2\sqrt{p(1-p)}$ is the traditional Bhattacharyya noise parameter for BSCs.

*C. Error Probability vs. $CB$ vs. $SB$*

Let $p_e = \mathsf{P}\left(X \neq \hat{X}_{\text{MAP}}(Y)\right)$ denote the error probability of the MAP decoder. The relationship between $p_e$ and the above noise measures $CB$ (or $\mathsf{CB}$) and $SB$ are stated by the following lemmas.

*Lemma 1:* For general BI-NSO channels and arbitrary input distributions, we have

$$\begin{aligned} 2p_e &\leq & CB &\leq 2\sqrt{p_e(1-p_e)}\\ 2p_e &\leq & SB &\leq 4p_e(1-p_e)\\ \text{and} & & SB \leq & CB &\leq \sqrt{SB}. \end{aligned}$$

*Lemma 2:* For any MI-SO channel with uniform input distribution, we have

$$2p_e \leq \sum_{x \in \mathbb{Z}_m\setminus\{0\}} CB(0 \to x).$$

If $p_e \leq 1/2$, then

$$\max_{x \in \mathbb{Z}_m\setminus\{0\}} CB(0 \to x) \leq 2\sqrt{p_e(1-p_e)}.$$

*Lemma 1* guarantees that the three statements: $p_e \to 0$, $CB \to 0$, and $SB \to 0$ are equivalent. *Lemma 2* guarantees $p_e \to 0$ is equivalent to the statement that $\forall x \in \mathbb{Z}_m\setminus\{0\}$, $CB(0 \to x) \to 0$. Detailed proofs of *Lemmata 1* and *2* are provided in APPENDIX II.

*D. The Equiprobable Graph Ensemble for LDPC Codes*

Throughout this paper, we consider only the equiprobable graph ensemble for LDPC codes [6], where each element corresponds to the Tanner graph of the parity check matrix. The construction of the equiprobable graph ensemble is as follows. Consider two finite sets of nodes: variable nodes and check nodes, in which each node is assigned a degree $\deg(v)$ (or $\deg(c)$) such that $\sum_{\text{all } v}\deg(v) = \sum_{\text{all } c}\deg(c) \triangleq n_e$. Assign $\deg(v)$ sockets to each variable node $v$ and index all variable sockets from 1 to $n_e$. Assign $\deg(c)$ sockets to each check node $c$ and index all check sockets from 1 to $n_e$. Let $\pi(\cdot)$ be an arbitrary permutation of the integers $1,\cdots,n_e$. Construct an edge connecting variable socket $i$ and check socket $j$ iff $i = \pi(j)$. The above procedure results in a bipartite graph and the equiprobable graph ensemble is constructed by letting $\pi(\cdot)$ being drawn from a uniform random permutation ensemble.

Based on this construction, we can define the edge-degree polynomials $\lambda(x) = \sum_k \lambda_k x^{k-1}$ and $\rho(x) = \sum_k \rho_k x^{k-1}$, where $\lambda_k$ (or $\rho_k$) is the percentage of the edges connecting to a variable (or check) node of degree $k$. This code ensemble can then be uniquely specified by the degree polynomials $\lambda$ and $\rho$ and the codeword length $n$, and is denoted by $\mathcal{C}^n(\lambda, \rho)$. We will sometimes use $\mathcal{C}(\lambda, \rho)$ to denote the asymptotic ensemble in the limit of large $n$. Further discussion of the asymptotic ensemble can be found in [6].

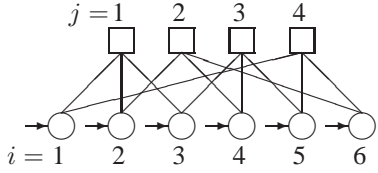

(a) The corresponding Tanner graph.

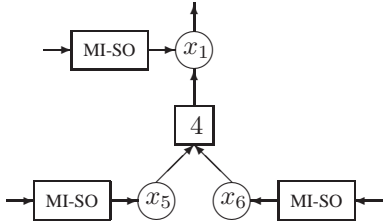

(b) The embedded support tree of depth $2l$, $l = 1$.

Fig. 3. Supporting tree of a regular (2,3) code.

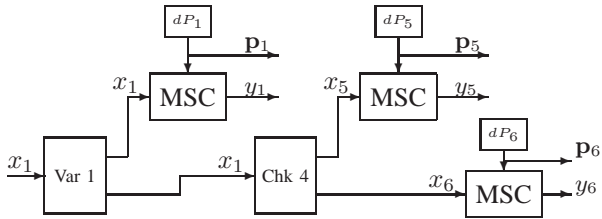

Fig. 4. LDPC codes with channels being decomposed as the probabilistic combinations of MSCs.

## III. THE SUPPORT TREE CHANNEL & EXISTING RESULTS

### A. The Support Tree Channel

Due to the inherent nature of message-exchanging during each iteration, the result of a BP decoder after $l$ iterations depends only on the neighbors (of the target variable node) within a distance of $2l$. With the assumption that there is no cycle of length less than $2l$ in the corresponding Tanner graph, which holds for sufficiently large codeword length $n$ in probability, the BP decoder on LDPC codes can be broken down into a tree structure of depth $2l$ as shown in [6] and demonstrated in Fig. 3, which considers the simplest case with target variable $x_1$ and $l = 1$. The arrows in Fig. 3(b) represent the message flow in the decoding process.

By the MSC decomposition argument, the tree structure can be viewed as in Fig. 4, which is a $\mathbb{Z}_m \mapsto ([0,1]^m \times \mathbb{Z}_m)^3$ vector channel. The arrows are now pointing in the opposite direction since they now represent the data flow during transmission. Due to its optimality when applied to a cycle-free inference network, the BP decoder is in essence an efficient version of a MAP decoder on the tree structure. Therefore, it is more convenient to focus on the behavior of *general* MAP decoders on this tree-like vector channel, instead of considering the message passing behavior of the BP decoder. It is worth emphasizing that throughout this paper, only independently and identically distributed channels are considered, and all $dP_1$, $dP_5$, $dP_6$ and their corresponding MSCs are independent.

Our target problem is to find bounds on the noise measures of the $\mathbb{Z}_m \mapsto ([0,1]^m \times \mathbb{Z}_m)^3$ vector output channel, given constraints of finite-dimensional noise measures on the constituent channel distribution $dP_i(\mathbf{p})$, $i = 1, 5, 6$. To simplify the problem further, we consider the variable node and the check node channels respectively as in Figs. 5(a) and 5(b), in which the new constituent $X_1 \mapsto (\mathbf{q}, Y')$ channel in Fig. 5(a) represents the entire $X_1 \mapsto (\mathbf{p}_5, Y_5) \times (\mathbf{p}_6, Y_6)$ vector channel in Fig. 5(b). Once the noise measure of the $X_1 \mapsto (\mathbf{p}_5, Y_5) \times (\mathbf{p}_6, Y_6)$ check node channel is bounded given the constraints on $X_5 \mapsto (\mathbf{p}_5, Y_5)$ and $X_6 \mapsto (\mathbf{p}_6, Y_6)$, this newly obtained bound for Fig. 5(b) can serve as a constraint on the constituent channel, $X_1 \mapsto (\mathbf{q}, Y')$, of the variable node channel in Fig. 5(a). When considering the behavior after infinitely many iterations, we can iteratively apply these check/variable node bounding techniques by switching the roles of "bounds on the vector output channel" and "constraints on the constituent channels" as in the aforementioned example. Given an initial constraint on the finite-dimensional noise measure of the constituent channels, whether the LDPC code is decodable can be determined by testing whether the noise measure converges to zero or is bounded away from zero as iterations proceed, which in turn gives us finite-dimensional lower/upper bounds on the decodable threshold.

For variable/check nodes with degrees $d > 3$, if we take the marginal approach (focusing on one input constituent channel while leaving other constituent channels fixed), all the effects of the fixed inputs can be grouped into a single input message. Therefore, it is as if we take the marginal approach on a variable/check node with degree equal to three. The analysis of nodes of degree one or two is trivial. As a result, throughout this paper, only nodes of degree $d = 3$ will be discussed in detail, and the variable/check node channels of interest are illustrated in Figs. 5(c) and 5(d) with inputs/outputs relabelled for easier reference.

### B. Existing Results on Binary LDPC Codes

For BI-SO channels, the best way to explain the existing results in [13], [14], [15], and [16] is using the idea of "transfer functions" and the convexity/concavity analysis. In this subsection, we will consider only the noise measure $CB$ for example, which will lead to the iterative upper bound in [14] and a new iterative lower bound. Similar arguments can be used to derive the results in [13] or in [15], [16], if we substitute either $SB$ or the conditional entropy for the noise measure $CB$.

*1) Check Nodes:* For a check node as in Fig. 5(d), the problem of finding an iterative upper/lower bound can be cast as an optimization problem as follows.

$$\max \text{ or } \min \quad CB_{out} = \int CB_{chk}(p,q) dP(p) dQ(q) \quad (7)$$

$$\text{subject to} \quad CB_{in,1} = \int CB_1(p) dP(p)$$

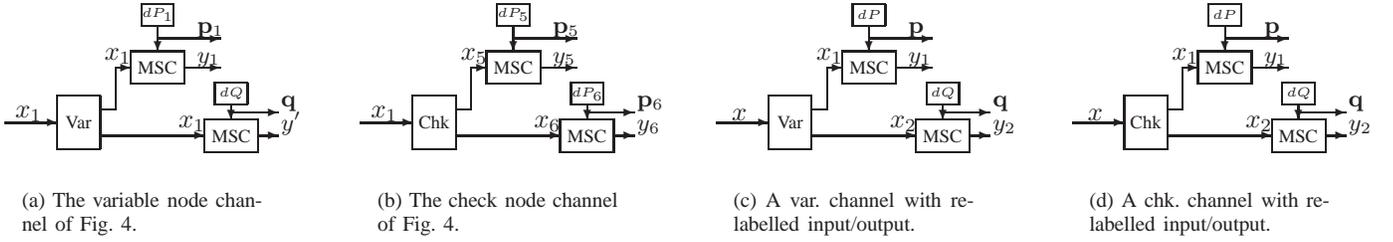

(a) The variable node channel of Fig. 4.

(b) The check node channel of Fig. 4.

(c) A var. channel with re-labelled input/output.

(d) A chk. channel with re-labelled input/output.

Fig. 5. Separate consideration of variable and check nodes.

$$CB_{in,2} = \int CB_2(q) dQ(q),$$

where

$$CB_{chk}(p,q) = 2\sqrt{p(1-p) + q(1-q) - 4p(1-p)q(1-q)} \quad (8)$$

$$CB_1(p) = 2\sqrt{p(1-p)}$$
$$CB_2(q) = 2\sqrt{q(1-q)}.$$

$CB_{chk}(p,q)$ denotes the value of $CB$ for the tree-like check node channel if both the constituent channels are BSCs with parameters $p$ and $q$ respectively, and $CB_1(p)$ and $CB_2(q)$ denote the $CB$ values of the constituent BSCs with parameters $p$ and $q$. Using some simple algebra, it can be shown that for fixed $p$ and $q$, the $X \mapsto Y_1 \times Y_2$ check channel is equivalent to a BSC with parameter $\pi = p(1-q) + (1-p)q$. Therefore $CB_{chk}(p,q) = 2\sqrt{\pi(1-\pi)} = 2\sqrt{p(1-p) + q(1-q) - 4p(1-p)q(1-q)}$, from which (8) follows. Furthermore, by omitting the input parameters $p$ and $q$, we can rewrite $CB_{chk}(p,q)$ in terms of $CB_1(p)$ and $CB_2(q)$ by

$$CB_{chk} = \sqrt{CB_1^2 + CB_2^2 - CB_1^2 CB_2^2}, \quad (9)$$

which is the $CB$-based "transfer function" of the check node. Since $CB_{chk}$ is a convex function of $CB_1$, this maximization/minimization problem is straightforward. The maximizing distribution $dP^*(p)$ is obtained by letting all probability weights concentrate on both extreme ends $p=0$ and $p=1/2$, that is

$$dP^*(p) = \begin{cases} 1 - CB_{in,1} & \text{if } p = 0 \\ CB_{in,1} & \text{if } p = 1/2 \\ 0 & \text{otherwise} \end{cases}.$$

Note: $dP^*$ is a probabilistic combination of a noise-free channel and a noisiest channel with output completely independent of the input, which corresponds to a BEC with erasure probability $\epsilon = CB_{in,1}$. One visualization of this maximizing solution can be obtained by fixing $CB_2$ and plotting all possible values of $(CB_1, CB_{chk})$ on a two dimensional plane, which form a convex curve. Connecting a string between both ends gives us the upper part of the convex hull of $(CB_1, CB_{chk})$'s. Therefore, the probabilistic combination of any $(CB_1, CB_{chk})$ lies within the convex hull and is upper bounded by the string. Since $dP^*(p)$ is a probabilistic weight such that the averaged $CB_1$ equals the constraint $CB_{in,1}$ and the averaged $CB_{chk}$ touches the string, $dP^*(p)$ must be a maximizing solution.

By Jensen's inequality, the minimizing distribution $dP^\dagger(p)$ is obtained by letting all probability weights concentrate on a single point with the same $CB_{in,1}$, that is

$$dP^\dagger(p) = \begin{cases} 1 & \text{if } CB_1(p) = 2\sqrt{p(1-p)} = CB_{in,1} \\ 0 & \text{otherwise} \end{cases}.$$

Note: $dP^\dagger$ corresponds to a BSC.

The same arguments can be applied to find $dQ^*$ and $dQ^\dagger$. By replacing both $dP$ and $dQ$ in (7) with the maximizing $dP^*$ and $dQ^*$, we prove that for general constituent BI-SO channels,

$$CB_{out} \leq CB_{in,1} + CB_{in,2} - CB_{in,1} CB_{in,2}.$$

By replacing both $dP$ and $dQ$ in (7) with the minimizing $dP^\dagger$ and $dQ^\dagger$, we also have

$$CB_{out} \geq \sqrt{CB_{in,1}^2 + CB_{in,2}^2 - CB_{in,1}^2 CB_{in,2}^2}.$$

By a straightforward extension to check nodes of higher degree $d_c \geq 3$, a similar upper bound can be obtained by replacing all $(d_c - 1)$ constituent channels[5] with BECs having the same values of $CB_{in,i}$. The resulting upper bound is

$$CB_{out} \leq 1 - \prod_{i=1}^{d_c-1} (1 - CB_{in,i}). \quad (10)$$

A similar lower bound can be obtained by replacing all $(d_c-1)$ constituent channels with BSCs having the same values of $CB_{in,i}$. The resulting lower bound is

$$CB_{out} \geq \sqrt{1 - \prod_{i=1}^{d_c-1} (1 - CB_{in,i}^2)}. \quad (11)$$

---

[5]A different bounding method is to represent a check node with $d_c > 3$ as a concatenation of $(d_c - 2)$ degree 3 check nodes and iteratively apply the bound derived for $d_c = 3$, the resulting bound of which is strictly looser than the bound constructed by direct replacement. For instance, [15] iteratively bounds the entropy by concatenating many degree 3 nodes, while [16] takes an approach similar to that in this paper and replaces all channels simultaneously, which results in tighter upper/lower bounds.

*2) Variable Nodes:* For a variable node as shown in Fig. 5(c), the problem of finding an iterative upper/lower bound can be cast as an optimization problem as follows.

$$\max \text{ or } \min \quad CB_{out} = \int CB_{var}(p,q)dP(p)dQ(q)$$
$$\text{subject to} \quad CB_{in,1} = \int CB_1(p)dP(p)$$
$$CB_{in,2} = \int CB_2(q)dQ(q),$$

where $CB_{var}(p,q)$ denotes the value of $CB$ for the tree-like variable node channel if both the constituent channels are BSCs with parameters $p$ and $q$ respectively. By the definition of $CB$ in (1), we have

$$\begin{aligned}&CB_{var}(p,q)\\ &= \sum_{x=0,1} P(X=x) \sum_{y_1,y_2 \in \{x,1-x\}^2} p(y_1,y_2|x)\sqrt{\frac{p(\bar{x}|y_1,y_2)}{p(x|y_1,y_2)}}\\ &= \sum_{x=0,1} \frac{1}{2}\left((1-p)(1-q)\sqrt{\frac{pq}{(1-p)(1-q)}} + p(1-q)\sqrt{\frac{(1-p)q}{p(1-q)}}\right.\\ &\quad \left. + (1-p)q\sqrt{\frac{p(1-q)}{(1-p)q}} + pq\sqrt{\frac{(1-p)(1-q)}{pq}}\right)\\ &= 4\sqrt{p(1-p)q(1-q)}, \end{aligned} \quad (12)$$

Omitting the input arguments $p$ and $q$, $CB_{var}$ can then be rewritten as

$$CB_{var} = CB_1 CB_2, \quad (13)$$

which is the $CB$-based "transfer function" of the variable node. Since $CB_{var}$ is a concave[6] function of $CB_1$, this maximization/minimization problem is straightforward and similar to the check node case. By Jensen's inequality, the maximizing distribution $dP^*(p)$ is obtained by letting all probability weights concentrate on a single point with the same $CB_{in,1}$, that is

$$dP^*(p) = \begin{cases} 1 & \text{if } CB_1(p) = 2\sqrt{p(1-p)} = CB_{in,1} \\ 0 & \text{otherwise} \end{cases},$$

which corresponds to a BSC. The minimizing distribution $dP^\dagger(p)$ is obtained by letting all probability weights concentrate on both extreme ends $p=0$ and $p=1/2$, that is

$$dP^\dagger(p) = \begin{cases} 1 - CB_{in,1} & \text{if } p = 0 \\ CB_{in,1} & \text{if } p = 1/2 \\ 0 & \text{otherwise} \end{cases},$$

which corresponds to a BEC. As a result, by replacing all constituent BI-SO channels with BSCs having the same values of $CB_{in,j}$, we obtain an upper bound for the variable node:

$$CB_{out} \leq \prod_{j=1}^{d_v-1} CB_{in,j}. \quad (14)$$

---
[6]Actually $CB_{var}$ is a linear function of $CB_1$. The reason we still view it as a concave function is to keep the argument reusable when we are considering other types of noise measures (eg., $SB$ and the conditional entropy).

By replacing all constituent channels with BECs having the same values of $CB_{in,j}$, we obtain a lower bound for the variable node:

$$CB_{out} \geq \prod_{j=1}^{d_v-1} CB_{in,j}. \quad (15)$$

*3) Combined Results:* Consider BI-SO channels and the irregular code ensemble with degree polynomials $\lambda$ and $\rho$. By combining (10) and (14) and averaging over the degree distributions, we have

$$CB^{(l+1)} \leq CB^{(0)}\lambda\left(1 - \rho\left(1 - CB^{(l)}\right)\right), \quad (16)$$

where $CB^{(l)}$ is the value of $CB$ after $l$ iterations, namely, the value of $CB$ for the support tree of depth $2l$. This is the result of Khandekar *et al.* in [14].

By combining (11) and (15) and averaging over $(\lambda, \rho)$, we have a new iterative lower bound.

*Theorem 1:* For BI-SO channels,

$$CB^{(l+1)} \geq CB^{(0)}\lambda\left(\sum_k \rho_k \sqrt{1 - \left(1 - (CB^{(l)})^2\right)^{k-1}}\right). \quad (17)$$

As stated in Section III-A, by checking whether $CB^{(l)}$ converges to zero by (16) or by (17), one can derive a lower/upper bound of the decodable threshold based on the $CB^{(0)}$ of the channel of interest. Closed form solutions for those thresholds can also be obtained following similar approaches as in [12], [16].

Similar arguments can be applied to other types of noise measures. In each iteration, replacing constituent channels of a check node with BECs/BSCs having the same value of $SB$, and replacing variable node constituent channels with BSCs/BECs having the same value of $SB$, we can reproduce the iterative upper/lower bound on $SB$ found in [13]. By considering the conditional entropy instead of $SB$, we can reproduce the iterative upper/lower bound on the mutual information found in [15], [16].

This paper will focus on developing new bounds or strengthening existing bounds for the cases in which the simple convexity/concavity analysis of the transfer function does not hold.

## IV. $\mathbb{Z}_m$ LDPC CODES

### A. Code Ensemble

The $\mathbb{Z}_m$-based LDPC code ensemble can be described as follows. The value of non-zero entries in the parity check matrix **H** are limited to one, and the random parity check matrix ensemble is identical to the ensemble of binary LDPC codes introduced in Section II-D. The only difference is that the parity check equation $\mathbf{Hx} = \mathbf{0}$ is now evaluated in $\mathbb{Z}_m$. A further deviation from the binary code ensemble is the $\mathsf{GF}(q)$-based code ensemble. Besides evaluating $\mathbf{Hx} = \mathbf{0}$ in $\mathsf{GF}(q)$, the non-zero entries in **H** are uniformly distributed between $\{1, 2, \cdots, q-1\}$. Further discussion of the $\mathbb{Z}_m$ and $\mathsf{GF}(q)$ LDPC code ensembles can be found in [18], [24].

## B. Iterative Bounds

*1) Variable Nodes:* As discussed in Section III-A, we focus on a variable node with degree $d_v = 3$ as in Fig. 5(c). We will first consider **p** and **q** being fixed (non-random) parameters and then extend our analysis to accommodate the random parameter generators $dP(\mathbf{p})$ and $dQ(\mathbf{q})$.

By grouping the outputs $Y_1$ and $Y_2$ into a 2-dimensional vector $\mathbf{Y} = (Y_1, Y_2)$, the variable node becomes a $\mathbb{Z}_m \mapsto \mathbb{Z}_m^2$ channel, and it can be verified by definition that it is still symmetric. By definition (4), the resulting $CB_{var}(0 \to x)$ for the vector output channel with uniformly distributed[7] $X$ becomes

$$CB_{var}(0 \to x)$$
$$= \sum_{\mathbf{y} \in \mathbb{Z}_m^2} \sqrt{(p_{y_1} q_{y_2})(p_{y_1-x} q_{y_2-x})}$$
$$= \left(\sum_{y_1 \in \mathbb{Z}_m} \sqrt{p_{y_1} p_{y_1-x}}\right) \cdot \left(\sum_{y_2 \in \mathbb{Z}_m} \sqrt{q_{y_2} q_{y_2-x}}\right)$$
$$= CB_{in,1}(0 \to x) \cdot CB_{in,2}(0 \to x).$$

A compact vector representation using the component-wise product "•" then becomes

$$\mathsf{CB}_{var} = \mathsf{CB}_1 \bullet \mathsf{CB}_2.$$

By iteratively applying the above inequality for variable nodes with $d_v \geq 3$, we have

$$\mathsf{CB}_{var} = \prod_{j=1}^{d_v-1} \mathsf{CB}_{in,j}, \tag{18}$$

where the $\prod$ represents the component-wise product. Consider general MI-SO constituent channels with random parameter generators $dP_j(\mathbf{p}^j)$, where $\mathbf{p}^j$ denotes the parameter vector for the $j$-th constituent channel and its distribution is denoted as $dP_j(\cdot)$. Since the parameter vectors $\mathbf{p}^j$ are independently distributed for different values of $j$, the probabilistic average of the product in (18) is the product of individual averages. This implies that (18) holds for general MI-SO channels as well.

*2) Check Nodes:* Consider a check node with degree $d_c = 3$, namely, two constituent MSCs with parameters **p** and **q**, as illustrated in Fig. 5(d). By definition, $CB_{chk}(0 \to x)$ for the $\mathbb{Z}_m \mapsto \mathbb{Z}_m^2$ channel is given as follows:

$$CB_{chk}(0 \to x)$$
$$= \sum_{w=0}^{m-1} \sqrt{\left(\sum_{y_1+y_2=w} p_{y_1} q_{y_2}\right)\left(\sum_{y_1+y_2=w+x} p_{y_1} q_{y_2}\right)}$$
$$= \sum_{w=0}^{m-1} \sqrt{\left(\sum_{y_1=0}^{m-1} p_{y_1} q_{w-y_1}\right)\left(\sum_{y_1=0}^{m-1} p_{y_1} q_{x+w-y_1}\right)}. \tag{19}$$

---
[7]By the linearity of the $\mathbb{Z}_m$ parity check code, the marginal distribution of any $X_i, \forall i \in \{1, \cdots, n\}$ is either uniform or concentrated on $\{0\}$. The latter case is of little interest since those bits can then be punctured without affecting the performance.

Each summand in (19) can be upper bounded by

$$\sqrt{\left(\sum_{y_1=0}^{m-1} p_{y_1} q_{w-y_1}\right)\left(\sum_{y_1=0}^{m-1} p_{y_1} q_{x+w-y_1}\right)}$$
$$= \sqrt{\left(\sum_{y=0}^{m-1} p_y q_{w-y}\right)\left(\sum_{z=0}^{m-1} p_z q_{x+w-z}\right)}$$
$$= \sqrt{\sum_{y=0}^{m-1}\sum_{z=0}^{m-1} p_y q_{w-y} p_z q_{x+w-z}}$$
$$\leq \sum_{y=0}^{m-1}\sum_{z=0}^{m-1} \sqrt{p_y p_z} \sqrt{q_{w-y} q_{x+w-z}}, \tag{20}$$

where the inequality follows from the fact that $\sqrt{\sum x_i} \leq \sum \sqrt{x_i}$ if $x_i \geq 0, \forall i$. By combining (19) and (20), we have

$$CB_{chk}(0 \to x)$$
$$\leq \sum_{w=0}^{m-1}\sum_{y=0}^{m-1}\sum_{z=0}^{m-1} \sqrt{p_y p_z} \sqrt{q_{w-y} q_{x+w-z}}$$
$$= \sum_{w'=0}^{m-1}\sum_{y=0}^{m-1}\sum_{z'=0}^{m-1} \sqrt{p_y p_{y+z'}} \sqrt{q_{w'} q_{w'+x-z'}} \tag{21}$$
$$= \sum_{z'=0}^{m-1} CB_{in,1}(0 \to z') CB_{in,2}(0 \to x - z'),$$

where (21) follows from the change of variables: $w' = w - y$ and $z' = z - y$. A compact vector representation using circular convolution "$\otimes$" then becomes

$$\mathsf{CB}_{chk} \leq \mathsf{CB}_{in,1} \otimes \mathsf{CB}_{in,2}. \tag{22}$$

By iteratively applying the above inequality and noting the monotonicity of the convolution operator (given all operands are component-wise non-negative), we have the following inequality for the cases $d_c > 3$:

$$\mathsf{CB}_{chk} \leq \bigotimes_{i=1}^{d_c-1} \mathsf{CB}_{in,i}. \tag{23}$$

Since the circular convolution is a summation of products and the $\mathbf{p}^j$ are independently distributed for different values of $j$, the probabilistic average of the circular convolution in (23) is the circular convolution of individual averages. This implies that (23) holds for general MI-SO channels as well.

*Note:* (22) is loose for the binary case ($m = 2$). For $m > 2$, there are many non-trivial cases in which (22) is tight. For example, suppose $m = 6$, $\mathbf{p}^1 = (0.5, 0.5, 0, 0, 0, 0)$, and $\mathbf{p}^2 = (0.5, 0, 0.5, 0, 0, 0)$. We have $\mathsf{CB}_{chk} = (1, 0.75, 0.5, 0.5, 0.5, 0.75)$, $\mathsf{CB}_{in,1} = (1, 0.5, 0, 0, 0, 0.5)$ and $\mathsf{CB}_{in,2} = (1, 0, 0.5, 0, 0.5, 0)$, which attains the equality in (22).

*3) Combined Results:* Consider general MI-SO channels and the irregular code ensemble with degree polynomials $\lambda$ and $\rho$. By combining (18) and (23) and averaging over the degree distributions, we have proved the following theorem.

*Theorem 2:* Let $\mathsf{CB}^{(l)}$ denote the value of CB for the support tree channel after $l$ iterations. Then we have

$$\mathsf{CB}^{(l+1)} \leq \mathsf{CB}^{(0)} \lambda\left(\rho\left(\mathsf{CB}^{(l)}\right)\right), \tag{24}$$

where the scalar products within $\lambda(\cdot)$ are replaced by component-wise products, and the scalar products within $\rho(\cdot)$ are replaced by circular convolutions.

For a $\mathbb{Z}_m$ code ensemble with degree polynomials $(\lambda, \rho)$, we can first fix an arbitrary[8] $\mathsf{CB}^*$, let $\mathsf{CB}^{(0)} = \mathsf{CB}^*$, and iteratively compute the upper bound by (24)[9]. Suppose that, for all $x \neq 0$, $\lim_{l \to \infty} CB^{(l)}(0 \to x) = 0$. By *Lemma 2*, any MI-SO channel with $\mathsf{CB} \leq \mathsf{CB}^*$ is guaranteed to be decodable by the BP algorithm when sufficiently long codes are used. Unlike the two-dimensional case, the thresholds determined herein for $m > 2$ cases do not admit straightforward closed form solutions due to the lack of ordering. Some further research is necessary to determine the closed form threshold "envelope" of the decodable CB vectors.

### C. Stability Conditions

The sufficient stability condition for $\mathbb{Z}_m$ LDPC codes can be obtained as a corollary to *Theorem 2*.

*Corollary 1 (Sufficient Stability Condition):* Consider any MI-SO channel with noise measure CB and any $\mathbb{Z}_m$ LDPC code ensemble with degree polynomials $(\lambda, \rho)$. If

$$\lambda_2 \rho'(1) CB(0 \to x) < 1, \quad \forall x \in \mathbb{Z}_m \setminus \{0\},$$

then this $\mathbb{Z}_m$ code is stable under the BP decoder. Namely, there exists $\epsilon > 0$ such that if after $l_0$ iterations, $\max_{x \in \mathbb{Z}_m \setminus \{0\}} CB^{(l_0)}(0 \to x) < \epsilon$, then $\lim_{l \to \infty} CB^{(l)}(0 \to x) = 0$ for all $x \neq 0$. (Or equivalently $\lim_{l \to \infty} p_e^{(l)} = 0$.) Furthermore, the convergence rate of $CB^{(l)}(0 \to x)$ is exponential or superexponential depending on whether $\lambda_2 > 0$ or $\lambda_2 = 0$.

*Proof:* Define $f_{CB}^{(l)} = \max_{x \in \mathbb{Z}_m \setminus \{0\}} CB^{(l)}(0 \to x)$. We prove the following equivalent statement that there exist $\epsilon, \delta > 0$ such that if $f_{CB}^{(l_0)} \in (0, \epsilon)$ for some $l_0$, we have

$$\frac{f_{CB}^{(l+1)}}{f_{CB}^{(l)}} < 1 - \delta, \quad \forall l > l_0. \tag{25}$$

Without loss of generality, we can assume $f_{CB}^{(l)} = \epsilon$. Using the fact that $CB^{(l)}(0 \to 0) = 1$ for all $l \in \mathbb{N}$, and the monotonicity of the convolution operator when all coordinates are positive, it can be shown that $\forall x \in \mathbb{Z}_m \setminus \{0\}$,

$$\left(CB^{(l)}\right)^{\otimes d_c - 1}(0 \to x) \leq ((d_c - 1)\epsilon + \mathcal{O}(\epsilon^2)),$$

where "$\otimes$" represents the convolution product. Similarly, for the component-wise product, one can show that

$$\left(CB^{(l)}\right)^{d_v - 1}(0 \to x) \leq \epsilon^{d_v - 1}.$$

[8]When uniform *a priori* distributions on $X$ are considered, any valid $\mathsf{CB}$ must satisfy the symmetric condition in (5) and that $CB(0 \to x) \in [0, 1]$, $\forall x \in \mathbb{Z}_m$.

[9]During the iterations, we may further strengthen (24) by $\mathsf{CB}^{(l+1)} \leq \min\left\{1, \mathsf{CB}^{(0)} \lambda\left(\rho\left(\mathsf{CB}^{(l)}\right)\right)\right\}$, since any valid $CB$ value is upper bounded by 1.

Using the above two inequalities and (24), we have

$$CB^{(l+1)}(0 \to x)$$
$$\leq CB^{(0)}(0 \to x) \sum_k \lambda_k \left(\sum_h \rho_h(h-1)\epsilon + \mathcal{O}(\epsilon^2)\right)^{k-1}$$
$$= CB^{(0)}(0 \to x) \lambda_2 \rho'(1) \epsilon + \mathcal{O}(\epsilon^2). \tag{26}$$

Since $\lambda_2 \rho'(1) CB(0 \to x) < 1, \forall x \in \mathbb{Z}_m \setminus \{0\}$, we can choose a $\delta > 0$ such that $1 - \delta > \lambda_2 \rho'(1) f_{CB}^{(0)}$. With a fixed choice of $\delta$, (25) is satisfied for sufficiently small $\epsilon$. (26) also shows that

$$\lim_{l \to \infty} \frac{f_{CB}^{(l+1)}}{f_{CB}^{(l)}} = f_{CB}^{(0)} \lambda_2 \rho'(1). \tag{27}$$

Hence the convergence rate is exponential or super exponential depending on whether $\lambda_2 = 0$. The proof is thus complete. ∎

A matching necessary stability condition can be proved as follows.

*Theorem 3 (Necessary Stability Condition):* Consider any MI-SO channel with noise measure CB and any $\mathbb{Z}_m$ LDPC code ensemble with degree polynomials $(\lambda, \rho)$. If

$$\exists x_0 \in \mathbb{Z}_m \setminus \{0\}, \text{ such that } \lambda_2 \rho'(1) CB(0 \to x_0) > 1,$$

then this $\mathbb{Z}_m$ code is not stable under the BP decoder. Namely, there exists $x_0 \in \mathbb{Z}_m \setminus \{0\}$ such that $\lim_{l \to \infty} CB^{(l)}(0 \to x_0) > 0$, or equivalently, $\lim_{l \to \infty} p_e^{(l)} > 0$.

A detailed proof using channel degradation argument similar to [6], [18] is provided in APPENDIX III.

We close this subsection by showing the stability results for $\mathsf{GF}(q)$ LDPC codes in [18] can be derived as a corollary to the above stability conditions for $\mathbb{Z}_m$ LDPC codes.

Consider a MI-SO channel with noise measure CB, and a $\mathsf{GF}(q)$-based LDPC code with degree polynomials $(\lambda, \rho)$, where $q$ is a prime number. The following stability conditions for $\mathsf{GF}(q)$ LDPC codes can be derived as direct corollaries to *Corollary 1* and *Theorem 3*, which were first presented in [18].

*Corollary 2 (Sufficient Stability Condition):* If

$$\lambda_2 \rho'(1) \frac{\sum_{x \in \mathsf{GF}(q) \setminus \{0\}} CB(0 \to x)}{q - 1} < 1,$$

then this $\mathsf{GF}(q)$ code is stable under the BP decoder.

*Corollary 3 (Necessary Stability Condition):* If

$$\lambda_2 \rho'(1) \frac{\sum_{x \in \mathsf{GF}(q) \setminus \{0\}} CB(0 \to x)}{q - 1} > 1,$$

then this $\mathsf{GF}(q)$ code is not stable under the BP decoder.

Since the stability conditions of the $\mathbb{Z}_m$ LDPC codes rely only on the pairwise error probability and the multiplication of the uniformly distributed edge weight $w \in \mathsf{GF}(q) \setminus \{0\}$ is equivalent to a uniform permutation of all non-zero entries, the stability conditions of a $\mathsf{GF}(q)$ code are equivalent to those of a $\mathbb{Z}_m$ code with the pairwise "error pattern" averaged over all non-zero entries. As a result, all results for $\mathbb{Z}_m$ codes involving only $CB(0 \to x)$ hold for $\mathsf{GF}(q)$ codes as well with each $CB(0 \to x)$ being replaced by the average $\frac{1}{q-1} \sum_{x=1}^{q-1} CB(0 \to x)$. The above corollaries then become simply the restatement of the stability conditions for $\mathbb{Z}_m$ LDPC codes.

*D. Applications*

We close this section with a non-comprehensive list of practical applications based on higher order LDPC codes and some corresponding references.

1) Improving the code performance [25]: By grouping two bits into a GF(4) symbol, the finite length performance of codes can be enhanced by the resulting higher dimensional codes.
2) Higher order coded modulation. Berkmann used higher order codes for coded modulation with the natural code-to-symbol mapping [26], [27]. By exhaustively searching for the optimal code-to-symbol mapper over all 8! possible mappers, our simulation shows that the asymptotic threshold of the regular $\mathbb{Z}_8$-based (3,9) code can be improved to within 0.63dB of the channel capacity of the 8PSK constellation with no additional cost. The simple structure of the (3,9) code and the better word-error probability make it an appealing alternative to turbo coded modulation or LDPC-coded bit-interleaved coded modulation (BICM).

    In addition to having lower decoding complexity, one argument that the BICM is favorable over $\mathbb{Z}_m$-based coded modulation is the higher cutoff rate of BICM [23], which leads to a common belief that it is easier to design good codes for BICM than for high order coded modulations. The higher cutoff rate of BICM can be better explained by the fact that the origin of the cutoff rate can be traced to the pairwise union bound (the Gallager bound) over different symbols, which results in the summation operation over all possible $CB(0 \to x)$ values in (6). Our stability conditions show that the "effective" value of $CB$ (in terms of the code performance) is the maximum of all $CB(0 \to x)$ rather than their summation. This result demonstrates that the cutoff rate, involving the sum rather than the maximum of $CB(0 \to x)$, is not a good benchmark between channels with different orders of alphabets. The above argument also gives a partial reasoning of the performance improvements after moving to a higher dimensional code in terms of the effective $CB$ value, since the max $CB(0 \to x)$ can be reduced by a good coded-alphabet to transmitting-symbol mapper, and a greater stability region can be obtained.

    *Note:* Another advantage of the higher order coded modulation over BICM is that a matched spectral null code can be concatenated as an inner code to improve the performance when in inter-symbol interference (ISI) environments [21].

3) Constructing mutual-information-achieving codes with non-uniform coded bit distribution by converting $\mathbb{Z}_m$ codes with uniform symbol distributions into non-uniformly distributed binary codes using "symbol mappers" [28]. Among the applications here are the following.
    - Constructing codes for cases in which the capacity-achieving *a priori* distribution is not uniform [28], [24].
    - Designing optimal superposition codes for broadcasting channels.
    - Designing codes for optical channels with cross talk [29].

Other references on higher order LDPC codes can be found in [30], [9]

## V. ITERATIVE BOUNDS ON BINARY CHANNELS

In this section, we will first show that the existing $CB$-based iterative bounds for BI-SO channels also hold for BI-NSO channels. Then we will strengthen the existing $CB$-based and $SB$-based bounds by providing a two-dimensional $(CB, SB)$-based iterative upper bound for BI-SO channels.

*A. $CB$-based Bounds on BI-NSO Channels*

The definition of $CB$ in (1) is applicable to BI-NSO channels with either uniform or non-uniform prior distributions. For the following, we will show that the inequalities (16) and (17) hold for BI-NSO channels as well by assuming a uniform prior distribution on $\mathbf{X} = \{0, 1\}$ and by adopting the reverse channel perspective. A uniform prior distribution is commonly assumed in all existing work on iterative bounds of LDPC code performance [13], [14], [15], [16], which can be justified by the perfect projection condition in [31].

*Theorem 4 ($CB$-based Bounds for BI-NSO Channels):* For the irregular LDPC code with degree polynomials $(\lambda, \rho)$, the iterative upper and lower bounds (16) and (17) hold for BI-NSO channels.

*Proof:* We will first focus on the simplest binary-input/binar-ouput non-symmetric channel, which is illustrated in Fig. 6(a) and is denoted as BNSC (in contrast to BSC). Since any BI-NSO channel can be regarded as the probabilistic combination of many BNSCs, our results for BNSC can then be generalized to arbitrary BI-NSO channels.

Any BNSC can be specified by two scalar parameters $p_{0\to 1}$ and $p_{1\to 0}$, where $p_{z\to z'}$ denotes the conditional probability $\mathsf{P}(Y = z'|X = z)$. $CB$ thus becomes

$$CB = \sqrt{p_{0\to 1}(1-p_{1\to 0})} + \sqrt{p_{1\to 0}(1-p_{0\to 1})}. \quad (28)$$

We can also represent this $X \mapsto Y$ BNSC from the reverse channel perspective $Y \mapsto X$ as in Fig. 6(b), such that

$$\begin{aligned} r_{0\to 1} &:= \frac{p_{1\to 0}}{1 - p_{0\to 1} + p_{1\to 0}} \\ r_{1\to 0} &:= \frac{p_{0\to 1}}{1 + p_{0\to 1} - p_{1\to 0}} \\ R(0) &:= \frac{1 - p_{0\to 1} + p_{1\to 0}}{2} \\ R(1) &:= \frac{1 + p_{0\to 1} - p_{1\to 0}}{2}, \end{aligned}$$

where $r_{z\to z'} := \mathsf{P}(X = z'|Y = z)$ and $R(y) = \mathsf{P}(Y = y)$. Then by definition, we have

$$\begin{aligned} CB &:= \mathsf{E}_{X,Y}\left\{\sqrt{\frac{p(\bar{X}|Y)}{p(X|Y)}}\right\} \\ &= R(0)2\sqrt{r_{0\to 1}(1-r_{0\to 1})} + R(1)2\sqrt{r_{1\to 0}(1-r_{1\to 0})} \\ &= R(0)CB(r_{0\to 1}) + R(1)CB(r_{1\to 0}), \end{aligned}$$

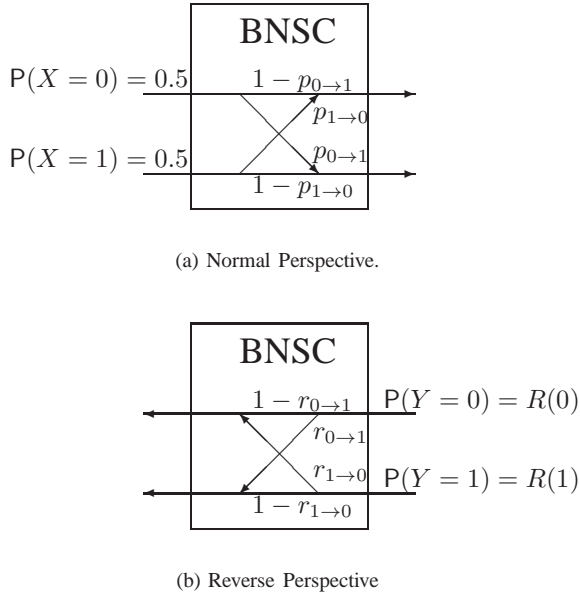

Fig. 6.  The probabilistic model of the BNSC.

in which $CB(p) \triangleq 2\sqrt{p(1-p)}$ computes the value of $CB$ for a BSC with crossover probability $p$. This representation separates the previous entangled expression of $CB$ in (28) so that it is as if there are two BSCs with parameters $r_{0\to 1}$ and $r_{1\to 0}$ respectively, and these two BSCs are probabilistically combined with coefficients $R(0)$ and $R(1)$. We use $(R(\cdot), \{r_{..}\})$ to represent this reverse channel perspective.

Consider the variable/check nodes of degree 3 with two constituent BNSCs in reverse form, namely, Ch1:$(R(\cdot), \{r_{..}\})$ and Ch2:$(S(\cdot), \{s_{..}\})$, such that

$$CB_{in,1} := \sum_{y_1 \in \{0,1\}} R(y_1) CB(r_{y_1 \to 0}),$$

$$CB_{in,2} := \sum_{y_2 \in \{0,1\}} S(y_2) CB(s_{y_2 \to 0}).$$

For a variable node of degree $d_v = 3$, by definition and after some simple algebra, we have

$$CB_{var} := \mathsf{E}_{X,Y} \left\{ \sqrt{\frac{p(\bar{X}|Y_1, Y_2)}{p(X|Y_1, Y_2)}} \right\}$$

$$= \mathsf{E}_{Y_1, Y_2} \left( \mathsf{E}_{X|Y_1, Y_2} \left( \sqrt{\frac{p(\bar{X}|Y_1, Y_2)}{p(X|Y_1, Y_2)}} \right) \right)$$

$$\stackrel{(a)}{=} \sum_{x \in \{0,1\}} \sum_{y_1 \in \{0,1\}} \sum_{y_2 \in \{0,1\}} \mathsf{P}(\mathbf{Y} = (y_1, y_2)|X_1 = X_2)$$
$$\cdot \sqrt{\mathsf{P}(X_1 = X_2 = x|\mathbf{Y} = (y_1, y_2), X_1 = X_2)}$$
$$\cdot \sqrt{\mathsf{P}(X_1 = X_2 = \bar{x}|\mathbf{Y} = (y_1, y_2), X_1 = X_2)}.$$

where $X_1$ and $X_2$ are the individual inputs of Channels 1 and 2. The last equality follows from the fact that the probability distribution with a variable node constraint $X = X_1 = X_2$ is identical to the conditional distribution by first assuming $X_1$ and $X_2$ are i.i.d. uniform Bernoulli distributions on $\{0, 1\}$ and then conditioning on the event $\{X_1 = X_2\}$. From the above equation, we can further simplify $CB_{var}$ as follows.

$CB_{var}$
$$= \sum_{y_1 \in \{0,1\}} \sum_{y_2 \in \{0,1\}} \frac{R(y_1) S(y_2) \mathsf{P}(X_1 = X_2 | \mathbf{Y} = (y_1, y_2))}{\mathsf{P}(X_1 = X_2)}$$
$$\cdot 2 \sqrt{\frac{\mathsf{P}(X_1 = X_2 = 0 | \mathbf{Y} = (y_1, y_2))}{\mathsf{P}(X_1 = X_2 | \mathbf{Y} = (y_1, y_2))}}$$
$$\cdot \sqrt{\frac{\mathsf{P}(X_1 = X_2 = 1 | \mathbf{Y} = (y_1, y_2))}{\mathsf{P}(X_1 = X_2 | \mathbf{Y} = (y_1, y_2))}}$$
$$= \sum_{y_1 \in \{0,1\}} \sum_{y_2 \in \{0,1\}} 4 R(y_1) S(y_2) \sqrt{r_{y_1 \to 0} s_{y_2 \to 0} r_{y_1 \to 1} s_{y_2 \to 1}}$$
$$= \left( \sum_{y_1 \in \{0,1\}} R(y_1) 2 \sqrt{r_{y_1 \to 0} r_{y_1 \to 1}} \right)$$
$$\cdot \left( \sum_{y_2 \in \{0,1\}} S(y_2) 2 \sqrt{s_{y_2 \to 0} s_{y_2 \to 1}} \right)$$
$$= CB_{in,1} \cdot CB_{in,2}, \quad (29)$$

By noting that (29) possesses the same form as in (13), all our previous analyses for variable nodes with BI-SO constituent channels hold for BI-NSO channels as well.

Consider a check node of degree $d_c = 3$, which is similar to Fig. 5(d) except that the constituent channels are now BNSCs. By definition, some simple algebra, and the observation that $X = X_1 + X_2$, we have

$CB_{chk}$
$$= \sum_{y_1 \in \{0,1\}} \sum_{y_2 \in \{0,1\}} \mathsf{P}(\mathbf{Y} = (y_1, y_2))$$
$$\cdot 2 \sqrt{\mathsf{P}(X_1 + X_2 = 0 | \mathbf{Y} = (y_1, y_2)) \mathsf{P}(X_1 + X_2 = 1 | \mathbf{Y} = (y_1, y_2))}$$
$$= \sum_{y_1 \in \{0,1\}} \sum_{y_2 \in \{0,1\}} R(y_1) S(y_2)$$
$$\cdot \sqrt{4 r_{y_1 \to 0} r_{y_1 \to 1} + 4 s_{y_2 \to 0} s_{y_2 \to 1} - 16 r_{y_1 \to 0} r_{y_1 \to 1} s_{y_2 \to 0} s_{y_2 \to 1}}$$
$$= \sum_{y_1 \in \{0,1\}} \sum_{y_2 \in \{0,1\}} R(y_1) S(y_2)$$
$$\cdot \sqrt{(CB(r_{y_1 \to 0}))^2 + (CB(s_{y_2 \to 0}))^2 - (CB(r_{y_1 \to 0}))^2 (CB(s_{y_2 \to 0}))^2}. \quad (30)$$

Note that (30) possesses the same form as in (7) and (9). Thus, each BNSC in (30) has the same effect as a BI-SO channel corresponding to a probabilistic combination of two BSCs with parameters $r_{0\to 1}$, $r_{1\to 0}$ (or $s_{0\to 1}$, $s_{1\to 0}$) and weights $R(0)$ and $R(1)$ (or $S(0)$ and $S(1)$). Since (10) and (11) hold for general BI-SO channels, they also hold for this particular combination of two BSCs, which in turn implies that they hold for BNSCs as well. By taking the probabilistic combination of many BNSCs, we have shown that (10) and (11) hold for general BI-NSO channels as well.

Since our previous analyses for both variable and check nodes (with BI-SO channels) hold for BI-NSO channels as well, we have proved *Theorem 4*. ∎

### B. A Two-Dimensional Upper Bound on BI-SO Channels

In this section, we develop a two-dimensional upper bound on the $(CB, SB)$ pair of a BI-SO channel, for which convexity/concavity analysis of the transfer function is not sufficient.

Similar to the one-dimensional results in Section III-B, we consider variable node and check nodes separately.

*1) Check Nodes:* Suppose the check node channel has two constituent BSCs with crossover probabilities $p \in [0, 1/2]$ and $q \in [0, 1/2]$ as shown in Fig. 5(d), where $p$ and $q$ have distributions $dP(p)$ and $dQ(q)$, respectively. Let $CB_{in,1}$ and $SB_{in,1}$ denote upper bounds on the values of $CB$ and $SB$ for the first constituent channel and let $CB_{in,2}$ and $SB_{in,2}$ denote corresponding upper bounds for the second constituent channel. We would like to develop an upper bound on the pair $(CB, SB)$ for the support tree channel. This iterative bounding problem thus becomes:

$$\max \quad CB_{out} = \int CB_{chk}(p,q) dP(p) dQ(q) \quad (31)$$
$$SB_{out} = \int SB_{chk}(p,q) dP(p) dQ(q)$$
$$\text{subject to} \quad \int 2\sqrt{p(1-p)} dP(p) \leq CB_{in,1}$$
$$\int 4p(1-p) dP(p) \leq SB_{in,1}$$
$$\int 2\sqrt{q(1-q)} dQ(q) \leq CB_{in,2}$$
$$\int 4q(1-q) dQ(q) \leq SB_{in,2},$$

where $CB_{chk}(p,q)$ is defined in (8) and

$$SB_{chk} := 4p(1-p) + 4q(1-q) - 4p(1-p)4q(1-q).$$

Using some simple algebra, we can show that the optimum value $SB_{out}^*$ satisfies $SB_{out}^* = SB_{in,1} + SB_{in,2} - SB_{in,1} SB_{in,2}$. The remaining problem reduces to the maximization of $CB_{out}$ subject to two input constraints on each of $dP$ and $dQ$. Solving this optimization problem, the maximizing $dP^*$ and $dQ^*$ can be expressed as follows.

$$dP^*(p) = \begin{cases} 1 - \frac{CB_{in,1}}{t} & \text{if } p = 0 \\ \frac{CB_{in,1}}{t} & \text{if } 2\sqrt{p(1-p)} = t \\ 0 & \text{otherwise} \end{cases},$$
$$\text{where} \quad t = \frac{SB_{in,1}}{CB_{in,1}}.$$

$dQ^*$ can be obtained by replacing $CB_{in,1}$, $SB_{in,1}$, and $p$ in the above equation with $CB_{in,2}$, $SB_{in,2}$, and $q$, respectively. A proof of the optimality of $dP^*$ and $dQ^*$ is given in APPENDIX IV.

By substituting all constituent BI-SO channels with channels of the same form as $dP^*$, we obtain an upper bound on $(CB, SB)$ in check node iterations as follows.

*Theorem 5 ($UB_{CB,SB}$ in Check Node Iterations):*
Suppose the check node degree is $d_c$ and the input $(CB, SB)$ pair is upper bounded by $(CB_{in}, SB_{in})$. Then the pair $(CB_{out}, SB_{out})$ of the check node iteration is bounded by

$$SB_{out} \leq 1 - (1 - SB_{in})^{d_c - 1}$$
$$CB_{out} \leq \sum_{i=1}^{d_c-1} \binom{d_c - 1}{i} \sqrt{1 - \left(1 - \left(\frac{SB_{in}}{CB_{in}}\right)^2\right)^i}$$
$$\left(1 - \frac{CB_{in}^2}{SB_{in}}\right)^{d_c - 1 - i} \left(\frac{CB_{in}^2}{SB_{in}}\right)^i . \quad (32)$$

*Corollary 4:* For the check node iteration of any $(\lambda, \rho)$ irregular LDPC codes, we have

$$SB_{out} \leq 1 - \rho(1 - SB_{in})$$
$$CB_{out} \leq \sum_k \rho_k \sum_{i=1}^{k-1} \binom{k-1}{i} \sqrt{1 - \left(1 - \left(\frac{SB_{in}}{CB_{in}}\right)^2\right)^i}$$
$$\left(1 - \frac{CB_{in}^2}{SB_{in}}\right)^{k-1-i} \left(\frac{CB_{in}^2}{SB_{in}}\right)^i.$$

*Note:* By incorporating the $SB_{in}$ constraint, the $CB$ bound (32) is now tight for both the BEC and BSC cases, which is a strict improvement over the $CB$-only bound (10). (The bound (10) is obtained by connecting the two ends of the $CB$-based transfer function curve and is tight for the BEC case but loose for the BSC case.)

*2) Variable Nodes:* We consider a variable node of degree $d_v = 3$. Given that the $(CB, SB)$ values of the constituent channels are upper bounded by $(CB_{in,1}, SB_{in,1})$ and $(CB_{in,2}, SB_{in,2})$, respectively, the iterative upper bounding problem becomes

$$\max \quad CB_{out} = \int CB_{var}(p,q) dP(p) dQ(q) \quad (33)$$
$$SB_{out} = \int SB_{var}(p,q) dP(p) dQ(q)$$
$$\text{subject to} \quad \int 2\sqrt{p(1-p)} dP(p) \leq CB_{in,1}$$
$$\int 4p(1-p) dP(p) \leq SB_{in,1}$$
$$\int 2\sqrt{q(1-q)} dQ(q) \leq CB_{in,2}$$
$$\int 4q(1-q) dQ(q) \leq SB_{in,2},$$

where $CB_{var}(p,q)$ is defined in (12) and

$$SB_{var}(p,q) := \frac{4p(1-p)4q(1-q)}{4p(1-p)(1 - 4q(1-q)) + 4q(1-q)}.$$

By some simple algebra, it can be shown that the optimum value $CB_{out}^*$ satisfies $CB_{out}^* = CB_{in,1} CB_{in,2}$. Unfortunately, for the remaining maximization problem on $SB_{out}$, the maximizing distribution $dP^*$ depends on both $(CB_{in,1}, SB_{in,1})$ and $(CB_{in,2}, SB_{in,2})$. The simple replacement of each constituent channel with a maximizing counterpart does not work this time. To circumvent this difficulty, we provide an upper bounding distribution $dP^{**}$ depending only on $(CB_{in,1}, SB_{in,1})$, such that the objective value of any feasible solutions $dP$ and $dQ$ is no larger than the objective value obtained from $dP^{**}$ and $dQ$. The distinction between the upper bounding distribution $dP^{**}$ and the maximizing distribution $dP^*$ is that $dP^{**}$ may not be feasible and thus may serve merely the bounding purpose.

For simplicity, we express $dP^{**}$ by dropping the subscript 1 in the vector constraint $(CB_{in,1}, SB_{in,1})$.

$$dP^{**}(p) = \begin{cases} (1-f_{SB})\frac{t}{t+CB_{in}} & \text{if } 2\sqrt{p(1-p)} = CB_{in} \\ f_{SB} & \text{if } 2\sqrt{p(1-p)} = \sqrt{SB_{in}} \\ (1-f_{SB})\frac{CB_{in}}{t+CB_{in}} & \text{if } 2\sqrt{p(1-p)} = t \\ 0 & \text{otherwise} \end{cases}, \quad (34)$$

where

$$t = \frac{SB_{in}}{CB_{in}}, \quad \text{which satisfies } CB \leq \sqrt{SB} \leq t,$$

$$f_{SB} = \begin{cases} 0 & \text{if } 2\sqrt{SB_{in}} - t + \sqrt{CB_{in}(2t-CB_{in})} \geq 0 \\ \frac{\eta(w^*)}{2t(t-CB_{in})^2} & \text{otherwise} \end{cases},$$

$$\eta(w) = w^3 - 2tw^2 + (t-CB_{in})^2 w, \quad (35)$$

$$w^* = \begin{cases} 2\sqrt{SB_{in}} & \text{if } \eta'(2\sqrt{SB_{in}}) \leq 0 \\ \frac{2t-\sqrt{4t^2-3(t-CB_{in})^2}}{3} & \text{otherwise} \end{cases}.$$

The upper bounding distribution $dQ^{**}$ for the second constituent channel can be obtained by symmetry. A derivation of $dP^{**}$ is included in APPENDIX V. It is worth noting that when there is no constraint on $CB_{in}$ (namely, when $CB_{in} = \sqrt{SB_{in}}$ by *Lemma 1*), $dP^{**}$ collapses to a BSC, which coincides with the $SB$-based bound in [13]. Hence the upper bound distribution $dP^{**}$ is a strict improvement over the existing $SB$-based bound.

Using this upper bounding distribution $dP^{**}$, an upper bound for $(CB, SB)$ for variable node iterations is given as follows.

*Theorem 6 ($UB_{CB,SB}$ in Variable Node Iterations):* Suppose the variable node degree is $d_v$, the input $(CB, SB)$ pair is upper bounded by $(CB_{in}, SB_{in})$, and the uncoded channel has noise measures $(CB_0, SB_0)$. Then, the output of the variable node iteration is upper bounded by

$$CB_{out} \leq CB_0(CB_{in})^{d_v-1},$$

and

$$SB_{out} \leq \Phi_{d_v-1}((CB_0, SB_0), (CB_{in}, SB_{in})),$$

where $\Phi_{d_v-1}$ computes the value of $SB$ for a variable node channel with one constituent $Ch_0$ channel and $(d_v - 1)$ constituent $Ch_{in}$ channels. Here, $Ch_{in}$ and $Ch_0$ are of the form of $dP^{**}$ and can be uniquely specified by $(CB_{in}, SB_{in})$ and $(CB_0, SB_0)$ respectively.

*Corollary 5:* For the variable node iteration of any $(\lambda, \rho)$ irregular LDPC codes, we have

$$\begin{aligned} CB_{out} &\leq CB_0 \lambda(CB_{in}) \\ SB_{out} &\leq \sum_k \rho_k \Phi_{k-1}((CB_0, SB_0), (CB_{in}, SB_{in})). \end{aligned}$$

An explicit expression for $\Phi_{k-1}$ involves a direct sum of various terms, the complexity of which grows at the order of $k^3$. A more practical, fast implementation is via the fast Fourier transform (FFT), which is similar to that used in density evolution. We first calculate the LLR message distribution $dP(m)$ for $Ch_0$ and $Ch_{in}$ from the upper bounding distribution $dP^{**}(p)$ in (34). Since the output LLR is the summation of input LLRs, the distribution of the output LLR is the convolution of the input LLRs, which can be calculated by FFT. At the end, we can use (3) to compute the corresponding output $SB$ value.

*3) Two-dimensional Iterative Upper Bound $UB_{CB,SB}$:* By combining the aforementioned upper bounds for the variable node and the check node iterations, we obtain a two-dimensional iterative upper bound $UB_{CB,SB}$. Since this two-dimensional bound is based on separate analysis of variable nodes and parity check nodes, it can be applied to any LDPC-like codes with graph-based ensembles without modification, including regular/irregular repeat accumulate (RA) codes [4], and joint-edge-distribution LDPC codes [32].

We omit the explicit expression for this two-dimensional bound since it is a direct concatenation of *Theorems 5* and *6*. By iteratively computing the upper bound $(CB^{(l)}, SB^{(l)})$ and testing whether it converges to $(0, 0)$, we can lower bound the decodable threshold for general BI-SO channels. The performance comparison of this procedure to existing results will be discussed in Section VII.

### C. Some Notes on Searching for High-Dimensional Bounds

Under the framework proposed in the previous sections, the problem of constructing iterative upper/lower bounds is equivalent to solving a corresponding optimization problem, within which the "variables" correspond to the probabilistic weight, $dP(p)$, of the corresponding BI-SO channel. Since most of the common noise measures of a BI-SO channel can be computed by the probabilistic average over the corresponding measures of the consitituent BSCs, both the constraints and the objective functions are generally linear with respect to $dP(p)$. The optimization of interest becomes a linear programming (LP) problem, of which the methods of finding optimal solutions ares well studied. Two notes about searching for upper/lower bounds are worth mentioning. First, when considering high-dimensional objective functions, the corresponding LP problem generally does not admit a uniform optimizer, as shown in our $(CB, SB)$ analysis in the previous sub-section, which hampers the use of simple channel replacement for bounding purpose. Second, the closed form solutions become more and more difficult to obtain when complicated constraints are applied, as demonstrated in (34). An alternative route is to use a commercial LP solver to numerically find bounds for each iteration. The closed form solution on the other hand is computationally much more efficient and provides better insight when compared to the numerical method. Since an iterative upper bound guarantees the minimum decodable threshold and is of higher importance from both the practical and theoretical perspectives, we use the two-dimensional upper bound to demonstrate this new framework and leave the two-dimensional lower bound for future research.

## VI. A One-Dimensional Non-Iterative Bound on BI-SO Channels

In this section, we construct a non-iterative upper bound, which is the best known bound that is tight for BSCs.

First we introduce some new notation. Let $p_e^{(l)}$ denote the bit error probability of the belief propagation after $l$ iterations. To distinguish between the types of BI-SO channels on which we are focusing, we append an argument $F_C$ to the end of $p_e^{(l)}$. That is, $p_e^{(l)}(F_C)$ denotes the bit error probability after $l$ iterations with the conditional distribution of the BI-SO channel being $F_C := \{f(y|x)\}$. In a similar fashion, we define $CB^{(l)}(F_C)$ as the Bhattacharyya noise parameter after $l$ iterations with the BI-SO channel being $F_C$, and $SB^{(l)}(F_C)$ is defined similarly. Following this definition, $CB(F_C) := CB^{(0)}(F_C)$ denotes the Bhattacharyya noise parameter of the uncoded BI-SO channel $F_C$. For simplicity, we use $F_{BSC,p}$ to denote the $F_C$ of a BSC with crossover probability $p$, and similarly we define $F_{BEC,\epsilon}$. Suppose for some $F_{BSC,\tilde{p}}$, the LDPC code ensemble is decodable. By the channel degradation argument in [6], one can show that all BI-SO channels $F_C$ with $p_e^{(0)}(F_C) \leq \tilde{p}$ are also decodable, a formal statement of which is as follows.

*Theorem 7 (The Channel Degradation Argument in [6]):* Suppose $F_C$ is a BI-SO channel and $F_{BSC,\tilde{p}}$ is a BSC. If $p_e(F_C) = p_e(F_{BSC,\tilde{p}})$, then for any $l \in \mathbb{N}$ and any irregular $(\lambda, \rho)$ LDPC codes,

$$p_e^{(l)}(F_{BSC,\tilde{p}}) \geq p_e^{(l)}(F_C).$$

The above inequality holds as well when subsituting $p_e^{(l)}(\cdot)$ by other common noise measures including $CB^{(l)}(\cdot)$, $SB^{(l)}(\cdot)$, and the conditional entropy.

This result, though being tight for BSCs, generally gives a very loose bound for other channels. We strengthen this result by providing a strictly tighter bound in the following theorems.

*Theorem 8:* Suppose $F_C$ is a BI-SO channel and $F_{BSC,\tilde{p}}$ is a BSC. If $SB(F_C) = SB(F_{BSC,\tilde{p}})$, then for any $l \in \mathbb{N}$ and any irregular $(\lambda, \rho)$ LDPC codes,

$$CB^{(l)}(F_{BSC,\tilde{p}}) \geq CB^{(l)}(F_C).$$

In *Theorem 8*, it is possible that $p_e^{(l)}(F_{BSC,\tilde{p}}) < p_e^{(l)}(F_C)$, which is different from the result using the channel degradation argument.

*Corollary 6:* If a $(\lambda, \rho)$ irregular LDPC code is decodable for a BSC with crossover probability $p^*$, then any BI-SO channel $F_C$ with $SB(F_C) \leq SB(F_{BSC,p^*}) = 4p^*(1 - p^*)$ is decodable under the same $(\lambda, \rho)$ code.

*Proof:* For any symmetric channel $F_C$ with $SB(F_C) \leq SB(F_{BSC,p^*})$, we consider a $F_{BSC,\tilde{p}}$ such that $SB(F_{BSC,\tilde{p}}) = SB(F_C)$. Since $F_{BSC,p^*}$ is physically degraded w.r.t. $F_{BSC,\tilde{p}}$, $F_{BSC,\tilde{p}}$ is also decodable, namely, $\lim_{l\to\infty} p_e^{(l)}(F_{BSC,\tilde{p}}) = 0$. By the relationship between $p_e$ and $CB$ in *Lemma 1* and by *Theorem 8*, we have

$$2p_e^{(l)}(F_C) \leq CB^{(l)}(F_C) \leq CB^{(l)}(F_{BSC,\tilde{p}})$$
$$\leq 2\sqrt{p_e^{(l)}(F_{BSC,\tilde{p}})} = o(1).$$

This completes the proof. ∎

Corollary 6 can be used as a tight one-dimensional upper bound, which is denoted by $UB_{SB}^*$.

A proof of *Theorem 8* is given in APPENDIX VI. We close this section by providing a lemma showing that *Theorem 8* is a strict improvement over *Theorem 7*, the channel degradation argument.

*Lemma 3:* Suppose $F_C$ is a BI-SO channel and $F_{BSC,\tilde{p}}$ is a BSC. If $p_e(F_C) = p_e(F_{BSC,\tilde{p}})$, then $SB(F_C) \leq SB(F_{BSC,\tilde{p}})$. Therefore, $\{F_C : SB(F_C) \leq SB(F_{BSC,\tilde{p}})\}$ is a super set of $\{F_C : p_e(F_C) \leq p_e(F_{BSC,\tilde{p}})\}$.

*Proof:* Let $dP_{F_C}(p)$ denote the probabilistic weight of the BSCs corresponding to the BI-SO channel $F_C$. Since $SB(F_C) = \int 4p(1-p) dP_{F_C}(p)$, $p_e(F_C) = \int p\, dP_{F_C}(p)$, and $4p(1-p)$ is a concave function of $p$, *Lemma 3* is a simple result of Jensen's inequality. ∎

## VII. Performance Comparisons

In this section, we compare the tightness of various lower bounds on the asymptotic decodable thresholds, obtained from the existing results and our results of Sections V-A, V-B, and VI.

Three existing results are included in TABLE I, including one based on the Bhattacharyya noise parameter [14], denoted as $UB_{CB}$, one on the soft bit value [13], denoted as $UB_{SB}$, and one on the conditional entropy $H(X|Y)$ [15], [16], denoted as $UB_{info}$. $UB_{CB,SB}$ denotes the two-dimensional $(CB, SB)$-based bound provided in Section V-B, and $UB_{SB}^*$ denotes the non-iterative tight bound given in Section VI. The DE column lists the asymptotic decodable thresholds obtained from density evolution [6]. In Section V-A, $UB_{CB}$ has been generalized for arbitrary BI-NSO channels. Therefore, the non-symmetric z-channel[10] is also included for comparison, in which the asymptotic threshold is obtained from the generalized density evolution method for BI-NSO channels [17].

As proved in Section V-B and evidenced in TABLE I, the two dimensional bound $UB_{CB,SB}$ provides strict improvement over $UB_{CB}$ and $UB_{SB}$. For channels that are neither BSC-like nor BEC-like, e.g. BiAWGNC and BiLC, the bound $UB_{info}$ found by Sutskover *et al.*, is tighter than $UB_{CB,SB}$ while the two dimensional $UB_{CB,SB}$ is tighter at both extreme ends. This phenomenon can be explained by the convexity/concavity analysis of the transfer functions. For $UB_{CB}$ the bounding inequality resides in the check node iteration, in which BECs attain the equality. Therefore, $UB_{CB}$ is the tightest when channels are BEC-like. For $UB_{SB}$, the bounding inequality resides in the variable node iteration, in which BSCs attain the equality, so $UB_{SB}$ is preferred for BSC-like channels. $UB_{CB,SB}$ thus has better performance in both extreme cases. On the other hand, $UB_{info}$ invokes bounding inequalities in both the variable node and the check node iterations. We observe that the absolute values of the curvatures of the transfer function is generally smaller when expressed in terms of the mutual information. Therefore, for the sake of insight, better predictability is obtained when the

---

[10]The z-channel is a BNSC such that $p_{1\to 0} > 0$ and $p_{0\to 1} = 0$.

Comparisons of lower bounds on decodable thresholds.

| | DE | $UB_{SB}$ | $UB_{CB}$ | $UB_{info}$ | $UB_{CB,SB}$ | $UB_{SB}^*$ |
|---|---|---|---|---|---|---|
| Decodable Thresholds | — | $SB^* \geq 0.2632$ | $CB^* \geq 0.4294$ | $h^* \geq 0.3644$ | — | $SB^* \geq 0.3068$ |
| BEC ($\epsilon^*$) | $\approx 0.4294$ | $\geq 0.2632$ | $\geq 0.4294$ | $\geq 0.3644$ | $\geq 0.4294$ | $\geq 0.3068$ |
| Rayleigh ($\sigma^*$) | $\approx 0.644$ | $\geq 0.5191$ | $\geq 0.6134$ | $\geq 0.6088$ | $\geq 0.6148$ | $\geq 0.5804$ |
| Z-channels ($p_{1 \to 0}^*$) | $\approx 0.2304$ | — | $\geq 0.1844$ | — | — | — |
| BiAWGNC ($\sigma^*$) | $\approx 0.8790$ | $\geq 0.7460$ | $\geq 0.7690$ | $\geq 0.8018$ | $\geq 0.7826$ | $\geq 0.8001$ |
| BiLC($\lambda^*$) | $\approx 0.65$ | $\geq 0.5610$ | $\geq 0.5221$ | $\geq 0.5864$ | $\geq 0.5670$ | $\geq 0.6146$ |
| BSC ($p^*$) | $\approx 0.0837$ | $\geq 0.0708$ | $\geq 0.0484$ | $\geq 0.0696$ | $\geq 0.0710$ | $\geq 0.0837$ |

TABLE I

COMPARISON OF LOWER BOUNDS DERIVED FROM FINITE-DIMENSIONAL ITERATIVE UPPER BOUNDS.

channel of interest is neither BSC nor BEC-like, e.g., the BiAWGN channel.

By *Lemma 1*, the feasible $(CB, SB)$ pairs satisfy $CB \geq SB$ and $(CB)^2 \leq SB$. By plotting general BI-SO channels according to their $(CB, SB)$ values, the set of decodable channels forms a "decodable region" and Fig. 7 demonstrates the decodable region of regular $(3, 6)$ codes. The decodable region is plotted with a thicker boundary using the channels considered in TABLE I. The density evolution method does not guarantee that all channels with $(CB, SB)$ inside the region are decodable. It is possible that two types of channels have the same $(CB, SB)$ values but one is decodable while the other is not.

The vertical line in Fig. 7 marked by $UB_{CB}$ represents the inner bound of the decodable $CB$ threshold [14]. The horizontal line marked by $UB_{SB}$ represents the inner bound of the decodable $SB$ threshold. Our results on the two-dimensional bound and the non-iterative tight bound greatly push the inner bounds of the decodable region toward its boundary (the curve marked $UB_{CB,SB}$ and the horizontal line marked $UB_{SB}^*$). These bounds guarantee that all BI-SO channels with $(CB, SB)$ within the inner bounds are decodable under belief propagation decoding. In Section V-A, we have shown that the vertical line $UB_{CB}$ holds as an inner bound even for BI-NSO channels.

It is worth noting that all bounds for binary-input channels in Sections V and VI, are obtained by simple channel replacement. Our proofs show that replacement of *any one* of the constituent channels will result in a new upper/lower bounding tree channel, and the replacement of *all* constituent channels gives us an upper/lower bound admitting *closed form solutions*, as those shown in the previous sections. In some situations, it is more advantageous to replace only part of the constituent channels, which results in tighter bounds at the expense of more complicated/channel-dependent solutions. For example, *Theorem 1* in [13] provides a different iterative upper bounding formula by substituting all $(d_v - 1)$ $Ch_{in}$ channels of a variable node with the corresponding maximizing BSCs while leaving the observation channel $Ch_0$ intact. The result is a channel-dependent iterative formula with tighter performance than the $UB_{SB}$ obtained by the replacement of all channels. The benefit of using selective channel replacement is also pointed out in [16].

A tight outer bound of the decodable region was proved by Burshtein *et al.* [13], illustrated by the horizontal line marked by $LB_{SB}$ in Fig. 7. Based on the mathematical symmetry between $CB$ and $SB$ in variable node and check node iterations, we conjecture the existence of a tight outer bound in terms of $CB$, which remains an open problem.

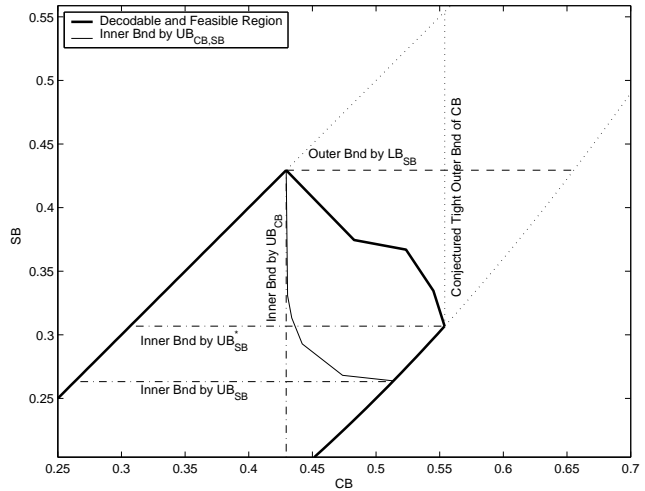

Fig. 7. The decodable region of the regular $(3, 6)$ code in the $(CB, SB)$ domain and some inner bounds of the decodable region.

## VIII. CONCLUSIONS

Finite dimensional bounds on the decodable thresholds find applications in both theoretical analysis and practical approximations. In this paper, we have developed a new iterative upper bound for $\mathbb{Z}_m$-based LDPC codes on MI-SO channels, which leads to a sufficient stability condition and provides insight into the analytical structure of general LDPC codes. Combined with a matching necessary stability condition proved herein, our stability condition pair can be used to derive the existing stability conditions for $GF(q)$-based LDPC codes.

Two new bounds for binary codes on BI-SO channels have also been constructed based on two types of noise measures, the Bhattacharyya noise parameter $CB$ and the soft bit value $SB$. These bounds push the existing inner bounds of the decodable region toward its boundary. An iteration-based bound for general memoryless BI-NSO channels, which finds applications in optical channels or magnetic storage channels, has also been derived.

Throughout this paper, a new framework enabling systematic searches for more finite-dimensional bounds has been provided, under which we have modelled the iterative bounding problem by considering its probabilistic decomposition. The performance discrepancy among various bounds can be explained by the tightness of different bounds during the variable node and the check node iterations. Besides the implied uniform good performance over all types of channels, these new finite dimensional bounds and the proposed framework provide a useful tool for studying the behavior of iterative decoding.

## APPENDIX I
## PROOF OF THE MSC DECOMPOSITION

*Proposition 1:* Any MI-SO channel, as illustrated in Fig. 8(a), can be converted to a probabilistic combination of many MSCs as in Fig. 8(c), the latter of which is equivalent to the original MI-SO channel from a detection point of view.

*Proof:* Without loss of generality, we assume $\mathbf{Y}$, the set of possible received values, is discrete. Since the original channel $\mathbb{Z}_m \mapsto \mathbf{Y}$ is symmetric, by *Definition 1*, there exists a bijective transformation $\mathcal{T}: \mathbf{Y} \mapsto \mathbf{Y}$ such that $\mathcal{T}^m(y) = y$ and $\forall x \in \mathbb{Z}_m, F(dy|0) = F(\mathcal{T}^x(dy)|x)$. Using $\mathcal{T}$, $\mathbf{Y}$ can be partitioned into many equivalence classes $\mathbf{Y}_j \subseteq \mathbf{Y}, j \in \mathbb{N}$, such that two elements $y_1$ and $y_2$ belong to the same class $\mathbf{Y}_j$ if and only if there exists an $x \in \mathbb{Z}_m$ such that $y_1 = \mathcal{T}^x(y_2)$. The original MI-SO channel can then be converted to an equivalent $x \mapsto (j, y)$ channel as in Fig. 8(b) such that $y \in \mathbf{Y}_j$ and $j \in \mathbb{N}$. Comparing Figs. 8(b) and (c), it remains to show that $dP(j|x)$ does not vary for different values of $x$, and for any $j$, there exists a MSC($\mathbf{p}$) with parameter $\mathbf{p}$ such that $F(dy|x, j)$ and the MSC($\mathbf{p}$) have the same distributions of the *a posteriori* probabilities regardless what type of the *a priori* distribution $P(X)$ is considered. Since the *a posteriori* probabilities are the sufficient statistics of any detection problem, the latter statement implies that Figs. 8(b) and (c) are equivalent channels from the detection point of view.

We first show that $dP(j|x)$ is not a function of $x$. By the construction of $\mathbf{Y}_j$ and by *Definition 1*, we have

$$\forall x \in \mathbb{Z}_m, \quad dP(j|x) = \sum_{y \in \mathbf{Y}_j} F(dy|x) = F(\mathbf{Y}_j|x)$$
$$= F(\mathcal{T}^{-x}(\mathbf{Y}_j)|0)$$
$$\stackrel{(a)}{=} F(\mathbf{Y}_j|0) = dP(j|0),$$

where (a) follows from the fact that $\mathbf{Y}_j$ is an equivalence class derived from the bijective transformation $\mathcal{T}$.

The second statement says that for every value $j$ of the *channel output* in Fig. 8(b), there exists a *side information* value $\mathbf{p}$ in Fig. 8(c) such that the posterior distribution $\mathsf{P}(x|\mathbf{Y}_j, j)$ from Fig. 8(b) is identical to the posterior distribution $\mathsf{P}(x|Y, \mathbf{p})$ from Fig. 8(c). To this end, we first let $y_0$ denote a "fixed" representative element of the non-empty[11] class $\mathbf{Y}_j$ from Fig. 8(b). We then define $y_i = \mathcal{T}^i(y_0)$. For Fig. 8(c), consider a MSC($\mathbf{p}$) with its parameter vector $\mathbf{p} = (p_0, \cdots, p_{m-1})$

[11] Without loss of generality, we may assume $\mathbf{Y}_j$ is non-empty, since an empty class is of little interest.

defined as follows:

$$p_i \propto \mathsf{P}(Y = y_i | X = 0, Y \in \mathbf{Y}_j), \forall i \in \mathbb{Z}_m, \quad (36)$$

where "$\propto$" means $p_i$ is proportional to the right-hand side while satisfying $\sum_{i \in \mathbb{Z}_m} p_i = 1$. The prior distribution of $X$ is the same for all models in Fig. 8 and is denoted by $\mathsf{P}(X = i) = \chi_i, \forall i \in \mathbb{Z}_m$. In the following proofs, it should be clear from the context which channel model in Fig. 8 we are considering.

Suppose $y_{i_0} \in \mathbf{Y}_j$ is received for Fig. 8(b). Then the *a posteriori* probabilities of $X = 0, 1, \cdots, m-1$ given $\mathbf{Y}_j$ and $j$ are proportional to $(\mathsf{P}(y_{i_0}|0, j)\chi_0, \mathsf{P}(y_{i_0}|1, j)\chi_1, \cdots, \mathsf{P}(y_{i_0}|m-1, j)\chi_{m-1})$. Again by the symmetry of the original channel, the *a posteriori* probabilities can be rewritten as

$$(\mathsf{P}(y_{i_0}|0, j)\chi_0, \mathsf{P}(\mathcal{T}^{-1}(y_{i_0})|0, j)\chi_1, \cdots,$$
$$\mathsf{P}(\mathcal{T}^{-(m-1)}(y_{i_0})|X = 0, j)\chi_{m-1})$$
$$= (\mathsf{P}(y_{i_0}|0, j)\chi_0, \mathsf{P}(y_{i_0-1}|0, j)\chi_1, \cdots,$$
$$\mathsf{P}(y_{i_0-(m-1)}|X = 0, j)\chi_{m-1})$$
$$\propto (p_{i_0}\chi_0, p_{i_0-1}\chi_1, p_{i_0-2}\chi_2, \cdots, p_{i_0-(m-1)}\chi_{m-1}).$$

By noting that the last equation also specifies the *a posteriori* probabilities given $Y = i_0$ from a MSC($\mathbf{p}$) with $\mathbf{p}$ specified in (36), it is proven that with $y_{i_0}$ and $i_0$ being the received values respectively, the partitioned channel $\mathbb{Z}_m \mapsto \mathbf{Y}_j$ in Fig. 8(b) has the same *a posteriori* probabilities as the MSC($\mathbf{p}$) in Fig. 8(c). To complete the proof that the partitioned channel $F(dy|x, j)$ has the same distribution of the *a posteriori* probabilities as the MSC($\mathbf{p}$), we need only to prove that the probability that $y_{i_0} \in \mathbf{Y}_j$ is received (in the partitioned channel $F(dy|x, j)$) is the same as the probability that $i_0 \in \mathbb{Z}_m$ is received (in the MSC($\mathbf{p}$)).

First consider the case in which $y_i \neq y_{i_0} \forall i \neq i_0$, and we then have $|\mathbf{Y}_j| = m$. Therefore

$$p_i = \mathsf{P}(Y = y_i | X = 0, Y \in \mathbf{Y}_j), \forall i \in \mathbb{Z}_m,$$

instead of being only proportional to the right-hand side. We then have

$$\mathsf{P}(Y = y_{i_0} | Y \in \mathbf{Y}_j)$$
$$= \sum_{i \in \mathbb{Z}_m} \mathsf{P}(Y = y_{i_0} | X = i, Y \in \mathbf{Y}_j)\chi_i$$
$$= \sum_{i \in \mathbb{Z}_m} \mathsf{P}(Y = y_{i_0-i} | X = 0, Y \in \mathbf{Y}_j)\chi_i$$
$$= \sum_{i \in \mathbb{Z}_m} p_{i_0-i}\chi_i$$
$$= P(\text{the output of the MSC}(\mathbf{p}) \text{ is } i_0).$$

For the case in which $\exists i \neq i_0$ such that $y_i = y_{i_0}$, unfortunately, $\mathsf{P}(Y = y_{i_0}|Y \in \mathbf{Y}_j)$ does not equal $\mathsf{P}(\text{the output of the MSC}(\mathbf{p}) \text{ is } i_0)$. However, it can be shown that all such $i$'s with $y_i = y_{i_0}$ will result in the same *a posteriori* probabilities. Furthermore, one can prove that

$$\mathsf{P}(Y = y_{i_0}|Y \in \mathbf{Y}_j)$$
$$= \mathsf{P}(\text{the output } i \text{ of the MSC}(\mathbf{p}) \text{ satisfies } y_i = y_{i_0}).$$

From the above discussion, the distribution of the *a posteriori* probabilities are the same for the partitioned channel $F(dy|x,j)$ and the MSC($\mathbf{p}$). The proof is thus complete. ∎

## APPENDIX II
### PROOF OF THE RELATIONSHIP AMONG $CB$, $SB$, AND $p_e$

Without loss of generality, we assume the conditional probability $\mathsf{P}(Y|X)$ is discrete, and all our derivations can be easily generalized to continuous/mixed situations.

*Proof of Lemma 1:* We use $q_{x,y} := \mathsf{P}(X=x, Y=y)$ to denote the joint probability of $X=x$ and $Y=y$. By definition, we have

$$p_e = \sum_{y \in \mathbf{Y}} \min(q_{0,y}, q_{1,y})$$
$$CB = 2 \sum_{y \in \mathbf{Y}} \sqrt{q_{0,y} q_{1,y}}$$
$$SB = 2 \sum_{y \in \mathbf{Y}} \frac{2 q_{0,y} q_{1,y}}{q_{0,y} + q_{1,y}}.$$

Since for any $x, y > 0$, $\min(x,y) \leq \frac{1}{\frac{1}{2}\left(\frac{1}{x}+\frac{1}{y}\right)} \leq \sqrt{xy}$, we immediately have $2p_e \leq SB \leq CB$. By Jensen's inequality and the concavity of the square root function, we can rewrite $CB$ as

$$\begin{aligned}CB &= \sum_{y \in \mathbf{Y}} (q_{0,y} + q_{1,y}) \sqrt{\frac{4 q_{0,y} q_{1,y}}{(q_{0,y}+q_{1,y})^2}} \\ &\leq \sqrt{\sum_{y \in \mathbf{Y}} (q_{0,y}+q_{1,y}) \frac{4 q_{0,y} q_{1,y}}{(q_{0,y}+q_{1,y})^2}} \\ &= \sqrt{SB}. \end{aligned} \quad (37)$$

Again by Jensen's inequality and the concavity of the polynomial $f(x) = 4x(1-x)$, we have

$SB$
$$= \sum_{y \in \mathbf{Y}} (q_{0,y}+q_{1,y}) 4 \frac{q_{0,y}}{q_{0,y}+q_{1,y}} \frac{q_{1,y}}{q_{0,y}+q_{1,y}}$$
$$= \sum_{y \in \mathbf{Y}} (q_{0,y}+q_{1,y}) \left[4 \frac{\min(q_{0,y}, q_{1,y})}{q_{0,y}+q_{1,y}} \left(1 - \frac{\min(q_{0,y}, q_{1,y})}{q_{0,y}+q_{1,y}}\right)\right]$$
$$\leq 4x(1-x)|_{x = \sum_{y \in \mathbf{Y}} (q_{0,y}+q_{1,y}) \frac{\min(q_{0,y}, q_{1,y})}{q_{0,y}+q_{1,y}}}$$
$$= 4 p_e (1 - p_e). \quad (38)$$

By (37), (38), and $2p_e \leq SB \leq CB$, the proof of *Lemma 1* is complete. ∎

*Proof of Lemma 2:* Define $p_{e,0 \leftrightarrow x}$ as the bit error probability of the MAP detector given that the input $X$ is uniformly distributed on $\{0, x\}$, namely,

$$p_{e, 0 \leftrightarrow x} := \mathsf{P}_{X \in_u \{0, x\}}\left(X \neq \hat{X}_{MAP}(Y)\right), \quad (39)$$

where $\mathsf{P}_{X \in_u A}(\cdot)$ denotes the probability assuming $X$ is evenly distributed on $A \subseteq \mathbb{Z}_m$. We note that $X \in_u \{0, x\}$ is equivalent to a binary-input channel with input alphabet $\{0, x\}$. Define $CB(0 \leftrightarrow x) = \frac{1}{2} CB(0 \to x) + \frac{1}{2} CB(x \to 0)$ as the $CB$ value of the binary channel $\{0, x\} \mapsto \mathbf{Y}$. Since $CB(0 \to x) = CB(x \to 0)$, we have $CB(0 \leftrightarrow x) = CB(0 \to x)$. By *Lemma 1*, we have

$$2 p_{e, 0 \leftrightarrow x} \leq CB(0 \to x) \leq 2 \sqrt{p_{e, 0 \leftrightarrow x}(1 - p_{e, 0 \leftrightarrow x})}.$$

From the above inequalities, the proof of *Lemma 2* can be completed by proving

$$\max_{x \in \mathbb{Z}_m \setminus \{0\}} \{p_{e, 0 \leftrightarrow x}\} \leq p_e \leq \sum_{x \in \mathbb{Z}_m \setminus \{0\}} p_{e, 0 \leftrightarrow x}. \quad (40)$$

We need only to prove the result for the MSC case, and the proof for general MI-SO channels then follows by taking the probabilistic average of the constituent MSCs. For an MSC with the parameter vector $\mathbf{p} = \{p_0, \cdots, p_{m-1}\}$, we have

$$p_{e, 0 \leftrightarrow x} = \frac{1}{2} \sum_{y=0}^{m-1} \min(p_y, p_{y+x}) \quad (41)$$
$$p_e = 1 - \max(p_0, p_1, \cdots, p_{m-1}).$$

Without loss of generality, we may assume $p_0$ is the maximum entry in $\mathbf{p}$ and $p_e = 1 - p_0$. Then for any $x \neq 0$, we can rewrite $p_{e, 0 \leftrightarrow x}$ as

$$\begin{aligned} p_{e, 0 \leftrightarrow x} &= \frac{1}{2}\left(p_x + \sum_{y=1}^{m-1} \min(p_y, p_{y+x})\right) \leq \frac{1}{2}\left(p_x + \sum_{y=1}^{m-1} p_y\right) \\ &\leq \frac{1}{2} \sum_{y=1}^{m-1} 2 p_y = 1 - p_0 = p_e, \end{aligned}$$

and the first half of Ineq. (40) is proved. Also by Eq. (41) and the assumption that $p_0$ is the maximal component of $\mathbf{p}$, we have

$$p_{e, 0 \leftrightarrow x} \geq \frac{p_x + p_{-x}}{2}, \forall x \in \mathbb{Z}_m \setminus \{0\}.$$

Summing over all possible $x \in \mathbb{Z}_m \setminus \{0\}$, the second half of (40) is also proved. ∎

## APPENDIX III
### NECESSARY STABILITY CONDITION FOR $\mathbb{Z}_m$ LDPC CODES

An $x$-erasure MSC can be defined by specifying its parameter vector $\mathbf{p}$ as $p_0 = 1/2$, $p_x = 1/2$ and $p_z = 0$, $\forall z \notin \{0, x\}$. Consider an $x$-erasure $\mathbb{Z}_m \mapsto \mathbb{Z}_m$ MSC, and suppose $0 \in \mathbb{Z}_m$ is received. From these specified conditional probabilities, $p_0, \cdots, p_{m-1}$, it is impossible for the receiver to determine whether the transmitting signal is $X = 0$ or $X = -x$ when a receiving value is $Y = 0$, which is as if we are facing a BEC, for which all information regarding $X = 0$ and $X = -x$ is erased. However, there is a fundamental difference between an $x$-erasure MSC and a BEC such that if we use $\mathbf{X} = \{0, -x'\}$ for transmission instead, the $x$-erasure MSC becomes a noiseless perfect channel assuming $x' \neq x$. In this section, we will use $\mathbf{e}_x$ to denote this particular parameter $\mathbf{p}$. An $x$-erasure decomposition lemma is given as follows.

*Lemma 4 ($x$-erasure Decomposition):* Consider any MI-SO channel with pairwise MAP error $p_{e, 0 \leftrightarrow x}$ defined in (39). This MI-SO channel can be written as a degraded channel of a probabilistic composition of two MSCs, of which the probabilistic weight $dQ^e(\mathbf{q})$ is defined as follows:

$$dQ^e(\mathbf{q}) = \begin{cases} 1 - 2p_{e,0\leftrightarrow x} & \text{if } \mathbf{q} = (1, 0, \cdots, 0) \\ 2p_{e,0\leftrightarrow x} & \text{if } \mathbf{q} = \mathbf{e}_x \\ 0 & \text{otherwise} \end{cases}.$$

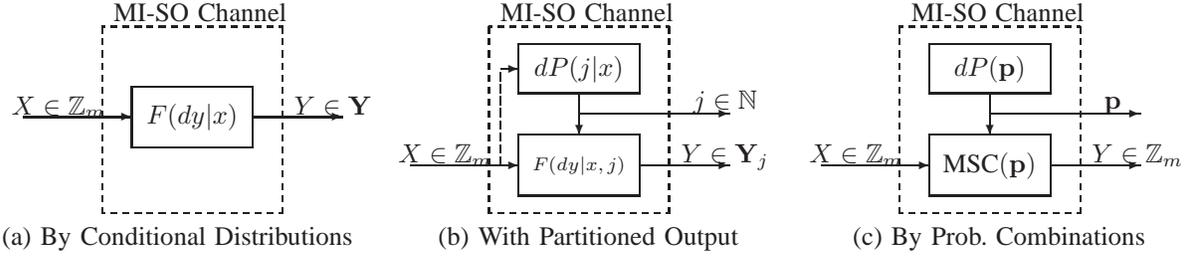

Fig. 8. Three equivalent representations of a MI-SO channel.

*Proof:* We need only to prove *Lemma 4* for an MSC with arbitrary parameter $\mathbf{p}$. By taking the average over $dP(\mathbf{p})$, the same result holds for general MI-SO channels.

We first note that $dQ^e(\cdot)$ can be viewed as a $\mathbb{Z}_m \mapsto (\{0,1\} \times \mathbb{Z}_m)$ channel, where the first output component is 1 iff $\mathbf{q} = \mathbf{e}_x$. Let $\mathbf{p}$ denote the parameter of the original MSC. We would like to show that there exists another channel $F$ such that after concatenating $dQ^e(\cdot)$ and $F$, we can reproduce the probability law $\mathbf{p}$ of the original MSC. To be more explicit, $F$ is a $(\{0,1\} \times \mathbb{Z}_m) \mapsto \mathbb{Z}_m$ channel such that the concatenation $\mathbb{Z}_m \mapsto (\{0,1\} \times \mathbb{Z}_m) \mapsto \mathbb{Z}_m$ becomes an MSC with parameter $\mathbf{p}$. We prove the existence of $F$ by explicitly specifying its probability law.

When the first component of the input of $F$ is given, say 0 or 1, let the remaining $\mathbb{Z}_m \mapsto \mathbb{Z}_m$ channel be an MSC with parameter $\mathbf{r}$ or with parameter $\mathbf{s}$ (depending on the first component being 0 or 1). Define, $\forall i \in \mathbb{Z}_m$,

$$r_i = \frac{p_i - \frac{1}{2}\min(p_i, p_{i+x}) - \frac{1}{2}\min(p_i, p_{i-x})}{1 - \sum_{z \in \mathbb{Z}_m} \min(p_z, p_{z+x})}$$

$$s_i = \frac{\min(p_i, p_{i+x})}{\sum_{z \in \mathbb{Z}_m} \min(p_z, p_{z+x})}.$$

It is easy to check that both $\mathbf{r}$ and $\mathbf{s}$ are valid parameter vectors.

It is also straightforward to check that the end-to-end $dQ^e \circ F$ channel is an MSC. By noting that $2p_{e,0\leftrightarrow x} = \sum_{z \in \mathbb{Z}_m} \min(p_z, p_{z+x})$, we can verify that the end-to-end channel has the same parameter $\mathbf{p}$ as the original MSC.
∎

Another necessary lemma is stated as follows.

*Lemma 5 (Monotonicity of $p_{e,0\leftrightarrow x}^{(l)}$):* Let $p_{e,0\leftrightarrow x}^{(l)}$ denote the pairwise error probability of the support tree channel of depth $2l$ (after $l$ iterations). Then $p_{e,0\leftrightarrow x}^{(l)}$ is non-increasing as a function of $l$. Furthermore, if $p_{e,0\leftrightarrow x}^{(l)} > 0$, then $p_{e,0\leftrightarrow x}^{(l+1)} > 0$.

*Proof:* As $l$ grows, the support tree gives more information by providing additional observations. As a result, the MAP error $p_{e,0\leftrightarrow x}^{(l)}$ is non-increasing as a function of $l$.

For the second statement, we break one iteration into its check node part and its variable node part. Since a check node channel is a degraded channel with respect to each of its constituent channels, we have $p_{e,0\leftrightarrow x}^{(l+\frac{1}{2})} \geq p_{e,0\leftrightarrow x}^{(l)} > 0$, where $p_{e,0\leftrightarrow x}^{(l+\frac{1}{2})}$ is the pairwise error probability of the support tree of depth $2l+1$ (after incorporating the check node). For variable nodes, by the equation $\mathsf{CB}_{var} = \mathsf{CB}_{in,1} \bullet \mathsf{CB}_{in,2}$, we have $p_{e,0\leftrightarrow x}^{(l+1)} = 0$ iff either $p_{e,0\leftrightarrow x}^{(0)} = 0$ or $p_{e,0\leftrightarrow x}^{(l+\frac{1}{2})} = 0$. Since both $p_{e,0\leftrightarrow x}^{(0)} \geq p_{e,0\leftrightarrow x}^{(l)} > 0$ and $p_{e,0\leftrightarrow x}^{(l+\frac{1}{2})} > 0$, it follows that $p_{e,0\leftrightarrow x}^{(l+1)} > 0$.
∎

*Proof of Theorem 3:* Suppose *Theorem 3* is false, namely, there exists a MI-SO channel such that $\lim_{l\to\infty} p_e^{(l)} = 0$ while there exists an $x_0 \in \mathbb{Z}_m \backslash \{0\}$ satisfying $\lambda_2 \rho'(1) CB(0 \to x_0) > 1$. By *Lemmata 1, 2,* and *5,* we have $\lim_{l\to\infty} p_{e,0\leftrightarrow x_0}^{(l)} = 0$ and $p_{e,0\leftrightarrow x_0}^{(l)} > 0, \forall l \in \mathbb{N}$, namely, $\{p_{e,0\leftrightarrow x_0}^{(l)}\}_{l \in \mathbb{N}}$ is a strictly positive sequence with limit 0. Therefore, for a sufficiently small $\epsilon > 0$, there exists an $l_0$ such that $p_{e,0\leftrightarrow x_0}^{(l_0)} < \epsilon$. Without loss of generality, we may assume $p_{e,0\leftrightarrow x_0}^{(l_0)} = \epsilon > 0$.

By *Lemma 4*, we can replace a supporting tree channel of depth $2l_0$ by a probabilistic combination of a perfect channel and an $x_0$-erasure channel with weights $(1 - 2\epsilon, 2\epsilon)$, which is denoted by $dQ^e(\mathbf{q})$. Similarly, for a supporting tree channel of depth $2(l_0 + \Delta l)$, we can replace each of its youngest subtrees of depth $2l_0$ by a $dQ^e(\mathbf{q})$ channel, so that after subsitition, the new channel becomes a supporting tree channel with depth $2\Delta l$ and all its youngest descendants are $dQ^e(\mathbf{q})$ channels. We then use $p_{e,0\leftrightarrow x_0}^{(l_0+\Delta l)}$ and $q_{e,0\leftrightarrow x_0}^{(\Delta l)}$ to denote the pairwise error probabilities of the original tree channel of depth $2(l_0 + \Delta l)$ and the new channel of depth $2\Delta l$ respectively. By the channel degradation argument, we have $p_{e,0\leftrightarrow x_0}^{(l_0+\Delta l)} \geq q_{e,0\leftrightarrow x_0}^{(\Delta l)}, \forall \Delta l \in \mathbb{N}$. It is worth noting that $p_{e,0\leftrightarrow x_0}^{(l_0)} = q_{e,0\leftrightarrow x_0}^{(0)}$.

For notational simplicity, we define the output of an $x_0$-erasure channel to be $+0$ if the output satisfies $Y = X + 0$. Similarly, an $x_0$-erasure channel outputs $+x_0$ if the output satisfies $Y = X + x_0$. For $\Delta l = 1$, we consider the new support tree channel of depth 2, namely, only one iteration of check node and variable node is considered. Readers are referred to Fig. 4 for illustration of a regular (2,3) LDPC code, in which both $dP_5$ and $dP_6$ should have the form of $dQ^e(\mathbf{q})$. Each check node constituent channel (CNCC), that is $dP_5$ or $dP_6$ in Fig. 4, can be either a noiseless perfect channel (with probability $1 - 2\epsilon$) or an $x_0$-erasure (with probability $2\epsilon$). If none of the check node constituent channels (CNCCs) is $x_0$-erasure, we can successfully decode the original input $X$ with no error. The cases that more than two CNCCs are $x_0$-erasure only happen with probability $\mathcal{O}(\epsilon^2)$, and are of less importance in the asymptotic analysis. Therefore we focus only on the case in which one and only one CNCC is $x_0$-erasure. Furthermore, since the input $X$ and the input of the individiual CNCC satisfy a parity check equation, the only sub-case in which there is no additional information (from the CNCCs) distinguishing $X = x_0$ from $X = 0$ is when the $x_0$-erasure CNCC has an output providing no information for detecting $-x_0$ from 0. We then have (42).

The inequality (a) in (42) follows from the fact that $+0$ is an output that the $x_0$-erasure CNCC cannot distinguish $-x_0$ from 0. Therefore, the event of misdetecting $X = 0$ by $X = x_0$ contains the case in which the CNCC outputs $+0$ given $X = 0$. Similarly, $+x_0$ is an output providing no information distinguishing 0 from $x_0$, which is thus contained by the event of misdetecting $X = x_0$ by $X = 0$. $2\epsilon\lambda_2\rho'(1)$ is the probability that one and only one CNCC is $x_0$-erasure, and $\frac{2\epsilon\lambda_2\rho'(1)}{2}$ corresponds to the probability for which the $x_0$-erasure channel outputs $+0$ (or $+x_0$). $m_x = \log\frac{P(Y=y|X=0)}{P(Y=y|X=x)}$ is the LLR between $X = 0$ and $X = x$, and $dP_x(\cdot)$ is the density of the initial LLR message $m_x$ given $X = 0$. Equality (b) in (42) follows from the fact that with the only $x_0$-erasure CNCC providing no information, misdectection happens when the original channel observation also provides an incorrect LLR message.

*Note:* If $x_0 \neq -x_0$ in $\mathbb{Z}_m$, the inequality becomes an equality. If $x_0 = -x_0$ in $\mathbb{Z}_m$, then $q_{e,0\leftrightarrow x_0}^{(1)}$ equals twice the right-hand side of the above expression.

By similar arguments, the second iteration gives

$$q_{e,0\leftrightarrow x_0}^{(2)} \geq \frac{1}{2}\epsilon\left(\lambda_2\rho'(1)\right)^2\left(\int_{m=-\infty}^0 (dP_{-x_0} \otimes dP_{x_0})(m)\right.$$
$$\left. + \int_{m=-\infty}^0 (dP_{x_0} \otimes dP_{-x_0})(m)\right) + \mathcal{O}(\epsilon^2),$$

and after $2\Delta l$ iterations we have

$$q_{e,0\leftrightarrow x_0}^{(2\Delta l)} \geq \epsilon\left(\lambda_2\rho'(1)\right)^{2\Delta l}\int_{m=-\infty}^0 (dP_{-x_0} \otimes dP_{x_0})^{\otimes \Delta l}(m)$$
$$+ \mathcal{O}(\epsilon^2).$$

It is easy to show that $(dP_{-x_0} \otimes dP_{x_0})$ is a symmetric distribution defined in [6], i.e., $dP(m) = e^m dP(-m)$, and its Bhattacharyya noise parameter is $(CB(0 \to x_0))^2$. Choose $\delta > 0$ such that $\lambda_2\rho'(1)(CB(0 \to x_0) - \delta) > 1$. By the tightness of the Bhattacharyya noise parameter, we can lower bound $q_{e,0\leftrightarrow x_0}^{(2\Delta l)}$ for sufficiently large $\Delta l$ by

$$q_{e,0\leftrightarrow x_0}^{(2\Delta l)} \geq \epsilon\left(\lambda_2\rho'(1)\right)^{2\Delta l}(CB(0 \to x_0) - \delta)^{2\Delta l} + \mathcal{O}(\epsilon^2).$$

Choose sufficiently large $\Delta l$ such that $(\lambda_2\rho'(1)(CB(0 \to x_0) - \delta))^{2\Delta l} > 2$ and sufficiently small $\epsilon$, we have $q_{e,0\leftrightarrow x_0}^{(2\Delta l)} \geq 2\epsilon + \mathcal{O}(\epsilon^2) > \epsilon$. By the channel degradation argument discussed earlier, we have

$$p_{e,0\leftrightarrow x_0}^{(l_0+2\Delta l)} \geq q_{e,0\leftrightarrow x_0}^{(2\Delta l)} > \epsilon = p_{e,0\leftrightarrow x_0}^{(l_0)},$$

which contradicts the monotonicity result in *Lemma 5*. Using this contradiction, the proof of *Theorem 3* is complete. ∎

## APPENDIX IV
### THE MAXIMIZING DISTRIBUTION FOR CHECK NODES WITH CONSTRAINTS ON $(CB_{in}, SB_{in})$

*Proof:* We take the approach of considering the marginal $dP$ first and assuming that $dQ$ is a point mass, i.e., $dQ$ concentrates all its probability on a fixed $q_0$. To simplify the notation, we let $a := 2\sqrt{p(1-p)}$ and $b := 2\sqrt{q_0(1-q_0)}$, and drop the subscript 1 in $(CB_{in,1}, SB_{in,1})$ to $(CB_{in}, SB_{in})$. The original problem (31) becomes a linear programming problem on $dP$:

$$\max \quad \zeta = \int_{a=0}^1 \sqrt{a^2(1-b^2)+b^2}\,dP(a)$$

$$\text{subject to} \int_{a=0}^1 dP(a) = 1$$
$$\int_{a=0}^1 a\,dP(a) \leq CB_{in}$$
$$\int_{a=0}^1 a^2\,dP(a) \leq SB_{in}$$
$$dP(a) \geq 0, \quad \forall a \in [0,1].$$

Note: From the BSC decomposition perspective, $dP(\cdot)$ denotes the probabilistic weight for different BSCs, which can be indexed by $p$ or by $a = 2\sqrt{p(1-p)}$ at one's will. Previously, $dP(\cdot)$ denotes the probabilistic weight for different BSCs indexed by $p$. Here the notation is slightly abused so that $dP(\cdot)$ also denotes the probabilistic weight for different BSCs indexed by $a = 2\sqrt{p(1-p)}$.

The corresponding dual problem is

$$\min_{y_0, y_1, y_2} \quad \xi := y_0 + y_1 CB_{in} + y_2 SB_{in}$$

$$\text{subject to } y_0 + ay_1 + a^2 y_2 \geq \sqrt{a^2(1-b^2)+b^2}, \quad \forall a \in [0,1]$$
$$y_1, y_2 \geq 0.$$

Let

$$t = \frac{SB_{in}}{CB_{in}},$$
$$y_0^* = b,$$
$$y_1^* = \frac{2}{t}\left(\frac{t^2(1-b^2)+2b^2}{2\sqrt{t^2(1-b^2)+b^2}} - b\right)$$
$$y_2^* = \frac{1}{t^2}\left(b - \frac{b^2}{\sqrt{t^2(1-b^2)+b^2}}\right).$$

It is easy to check that both $y_1^*, y_2^* \geq 0$. By *Lemma 6* (stated at the end of this proof), $\mathbf{y}^* \triangleq (y_0^*, y_1^*, y_2^*)$ is a feasible solution. Let

$$dP^*(a) = \begin{cases} 1 - \frac{CB_{in}}{t} & \text{if } a = 0 \\ \frac{CB_{in}}{t} & \text{if } a = t \\ 0 & \text{otherwise} \end{cases}.$$

It can be verified that the duality gap between the two feasible solutions $\mathbf{y}^*$ and $dP^*$ is zero. By the weak duality theorem of linear programming, $dP^*$ is the maximizing distribution when $dQ$ concentrates on $q_0$. Since $dP^*$ does not depend on $b$ (and thus does not depend on $q_0$), $dP^*$ is the universal maximizer for general $dQ$. ∎

*Lemma 6:* $y_0^* + ay_1^* + a^2 y_2^* \geq \sqrt{a^2(1-b^2)+b^2}$ for all $a, b \in [0,1]$.

*Proof:* Let $f(a) := y_0^* + ay_1^* + a^2 y_2^* - \sqrt{a^2(1-b^2)+b^2}$ be a function of $a$ while $b$ and $t$ are fixed parameters. We first

$$q_{e,0\leftrightarrow x_0}^{(1)} = \mathsf{P}(X=0)\mathsf{P}\left(\hat{X}_{MAP}=x_0|X=0\right) + \mathsf{P}(X=x_0)\mathsf{P}\left(\hat{X}_{MAP}=0|X=x_0\right) \quad (42)$$

$$\overset{(a)}{\geq} \frac{1}{2}\mathsf{P}(\text{one and only one CNCC is } x_0\text{-erasure and that channel outputs } +0|X=0)$$
$$\cdot \mathsf{P}\left(\hat{X}_{MAP}=x_0|\text{one and only one CNCC is } x_0\text{-erasure and outputs } +0, X=0\right)$$
$$+ \frac{1}{2}\mathsf{P}(\text{one and only one CNCC is } x_0\text{-erasure and outputs } +x_0|X=x_0)$$
$$\cdot \mathsf{P}\left(\hat{X}_{MAP}=0|\text{one and only one CNCC is } x_0\text{-erasure and outputs } +x_0, X=x_0\right)$$
$$+ \mathcal{O}(\epsilon^2)$$
$$\overset{(b)}{=} \frac{1}{2}\left(\frac{2\epsilon\lambda_2\rho'(1)}{2}\right)\int_{m_{x_0}=-\infty}^{0} dP_{x_0}(m_{x_0}) + \frac{1}{2}\left(\frac{2\epsilon\lambda_2\rho'(1)}{2}\right)\int_{m_{-x_0}=-\infty}^{0} dP_{-x_0}(m_{-x_0}) + \mathcal{O}(\epsilon^2),$$

note that

$$\frac{d^3 f}{da^3}(a) = \frac{3ab^2(1-b^2)^2}{\left(\sqrt{a^2(1-b^2)+b^2}\right)^5} \geq 0, \quad (43)$$

$$f(0) = 0,$$
$$f(t) = 0,$$

and

$$\left.\frac{df}{da}\right|_{a=t} = 0.$$

By simple calculus, the conclusion $f(a) \geq 0, \forall a \in [0,1]$ can be obtained in different ways, one of which is demonstrated as follows.

We first show by contradiction that there exists no $a_0$ other than 0 and $t$ such that $f(a_0) = 0$. Suppose there exists an $a_0 \in (0,1)$, such that $a_0 \neq 0, t$, and $f(a_0) = 0$. Since $f(0) = f(t) = 0$, by the mean value theorem (MVT), there exist $a_1 < a_2 \in (0, \max(a_0, t))$ such that $a_1, a_2 \neq t$, and $f'(a_1) = f'(a_2) = 0$. Since $f'(t) = 0$, by the MVT, there exist $a_3 < a_4 \in (\min(a_1, t), \max(a_2, t))$ such that $f^{(2)}(a_3) = f^{(2)}(a_4) = 0$. Again by the MVT, $\exists a_5 \in (a_3, a_4)$, such that $f^{(3)}(a_5) = 0$. By (43), the only possibility that such an $a_5 \in (0,1)$ exists is when $f^{(3)}(\cdot) = 0$ is a zero-function, which contradicts the assumption that the minimal number of distinct roots is no less than 3 (with values 0, $t$, and $a_0$).

Since there exists no $a_0$ other than 0 and $t$ such that $f(a_0) = 0$, the only case that $\exists a \in [0,1], f(a) < 0$ is when one of the following statements holds: (i) $f(a) < 0$ for all $a \in (0, t)$, or (ii) $f(a) < 0$ for all $a \in (t, 1)$. Suppose (i) holds. A contradiction can be obtained by consecutively applying the MVT as follows. Since $f(0) = f(t) = 0$, $\exists a_1 \in (0, t)$ such that $f'(a_1) = f'(t) = 0$. Therefore, $\exists a_2 \in (a_1, t)$ such that $f''(a_2) = 0$ and $f(a_2) < 0$. Since $f(t) = 0$, $\exists a_3 \in (a_2, t)$ such that $f'(a_3) > 0$. Since $f'(t) = 0$, $\exists a_4 \in (a_3, t)$ such that $f''(a_4) < 0$. Therefore, $\exists a_5 \in (a_2, a_4)$ such that $f^{(3)}(a_5) < 0$, which contradicts (43).

The remaining case is when (ii) holds and $f(a) > 0, \forall a \in (0, t)$. Since $f(t) = 0$, by the MVT, $\exists a_1 \in (0, t), a_2 \in (t, 1]$ such that $f'(a_1) < 0$ and $f'(a_2) < 0$. Since $f'(t) = 0$, $\exists a_3 \in (a_1, t), a_4 \in (t, a_2)$ such that $f''(a_3) > 0$ and $f''(a_4) < 0$. Therefore $\exists a_5 \in (a_3, a_4)$ such that $f^{(3)}(a_5) < 0$, which contradicts (43). From the above discussion, the proof is complete. ∎

## APPENDIX V
### THE UPPER BOUNDING DISTRIBUTION FOR VARIABLE NODES WITH CONSTRAINTS ON $(CB_{in}, SB_{in})$

We take the approach of assuming $dQ$ concentrates on a fixed $q_0$. Let $a := 2\sqrt{p(1-p)}$ and $b := 2\sqrt{q_0(1-q_0)}$ and drop the subscript 1 in $(CB_{in,1}, SB_{in,1})$ to write $(CB_{in}, SB_{in})$. The original problem (33) becomes a linear programming problem with the primal and dual representations as follows.

$$\max \quad \zeta := \int \frac{a^2 b^2}{a^2(1-b^2)+b^2} dP(a)$$
$$\text{subject to} \quad \int dP(a) = 1$$
$$\int a\, dP(a) \leq CB_{in}$$
$$\int a^2\, dP(a) \leq SB_{in}$$
$$dP(a) \geq 0, \quad \forall a \in [0,1]$$

$$\min_{y_0, y_1, y_2} \quad \xi := y_0 + y_1 CB_{in} + y_2 SB_{in}$$
$$\text{subject to} \quad y_0 + ay_1 + a^2 y_2 \geq \frac{a^2 b^2}{a^2(1-b^2)+b^2}, \quad \forall a \in [0,1]$$
$$y_1, y_2 \geq 0.$$

For convenience, we define $t := SB_{in}/CB_{in}$ and $r_b(a) := \frac{a^2 b^2}{a^2(1-b^2)+b^2}$.

Unlike the check node channel case, this time the optimal primal solution $dP^*(a)$ depends on the value of $b$, and different values of $b$ will lead to different closed form solutions of the optimizer $dP^*(a)$. The effect of different $b$'s can be summarized as three different cases in which $b$ belongs to one of the following three intervals, $\left[0, \sqrt{\frac{CB_{in}^2}{1+CB_{in}^2}}\right]$, $\left[\sqrt{\frac{CB_{in}^2}{1+CB_{in}^2}}, \sqrt{\frac{t^2}{1+t^2}}\right]$, and $\left[\sqrt{\frac{t^2}{1+t^2}}, 1\right]$, respectively.

*Proposition 2:* If $b \in \left[0, \sqrt{\frac{CB_{in}^2}{1+CB_{in}^2}}\right]$, the maximizing

$dP^*(a)$ and the optimum values are as follows.

$$dP^*(a) = \begin{cases} 1 & \text{if } a = CB_{in} \\ 0 & \text{otherwise} \end{cases}, \quad (44)$$

$$\zeta^* = \xi^* = \frac{CB_{in}^2 b^2}{CB_{in}^2(1-b^2) + b^2}.$$

*Proof:* It is easy to check that the specified $dP(a)$ is feasible. We then note that $r_b(0) = 0$, and $\left(\frac{1}{2}\sqrt{\frac{b^2}{1-b^2}}\right) \cdot a$ is the only tangent line of $r_b(a)$ passing through the origin (with the contact point $a_t = \sqrt{\frac{b^2}{1-b^2}}$). Furthermore, when $a \geq \sqrt{\frac{b^2}{1-b^2}}$, we have $\frac{d^2 r_b}{da^2} = 2b^4 \frac{b^2 - 3a^2(1-b^2)}{(a^2(1-b^2)+b^2)^3} \leq 0$ and $r_b(a)$ is thus a concave function in the interval $[\sqrt{\frac{b^2}{1-b^2}}, 1]$. From the above observations,

$$u_b(a) = \begin{cases} \left(\frac{1}{2}\sqrt{\frac{b^2}{1-b^2}}\right) \cdot a & \text{if } a \in [0, \sqrt{\frac{b^2}{1-b^2}}] \\ r_b(a) & \text{if } a \in [\sqrt{\frac{b^2}{1-b^2}}, 1] \end{cases} \quad (45)$$

is the convex hull of $r_b(a)$. By Jensen's inequality,

$$\int r_b(a) dP^*(a) \leq \int u_b(a) dP^*(a)$$
$$\leq u_b\left(\int a dP^*(a)\right) = \frac{CB_{in}^2 b^2}{CB_{in}^2(1-b^2) + b^2}. \quad (46)$$

Since $dP^*(a)$ in (44) achieves the upper bound in (46), it is indeed the maximizing distribution. ∎

*Proposition 3:* If $b \in \left[\sqrt{\frac{CB_{in}^2}{1+CB_{in}^2}}, \sqrt{\frac{t^2}{1+t^2}}\right]$, the maximizing $dP^*(a)$ and the optimum values are as follows.

$$dP^*(a) = \begin{cases} CB_{in}\sqrt{\frac{1-b^2}{b^2}} & \text{if } a = \sqrt{\frac{b^2}{1-b^2}} \\ 1 - CB_{in}\sqrt{\frac{1-b^2}{b^2}} & \text{if } a = 0 \\ 0 & \text{otherwise} \end{cases}, \quad (47)$$

$$\zeta^* = \xi^* = \frac{CB_{in}}{2}\sqrt{\frac{b^2}{1-b^2}}$$

*Proof:* It is easy to check that the specified $dP^*(a)$ is feasible. By again invoking Jensen's inequality on $u_b(a)$ defined in (45), we have

$$\int r_b(a) dP^*(a) \leq \int u_b(a) dP^*(a)$$
$$\leq u_b\left(\int a dP^*(a)\right) = \frac{CB_{in}}{2}\sqrt{\frac{b^2}{1-b^2}}. \quad (48)$$

Since $dP^*(a)$ in (47) achieves the upper bound in (48), it is indeed the maximizing distribution. ∎

*Proposition 4:* If $b \in \left[\sqrt{\frac{t^2}{1+t^2}}, 1\right]$, the maximizing $dP^*(a)$ and the optimum values are as follows.

$$dP^*(a) = \begin{cases} \frac{CB_{in}^2}{SB_{in}} & \text{if } a = t \\ 1 - \frac{CB_{in}^2}{SB_{in}} & \text{if } a = 0 \\ 0 & \text{otherwise} \end{cases},$$

$$\zeta^* = \xi^* = \frac{SB_{in} b^2}{t^2(1-b^2) + b^2}$$

*Proof:* It is easy to check that the specified $dP^*(a)$ is feasible. By choosing

$$y_0^* = 0$$
$$y_1^* = \frac{2t^3 b^2(1-b^2)}{(t^2(1-b^2)+b^2)^2} \geq 0$$
$$y_2^* = \frac{b^2(b^2 - t^2(1-b^2))}{(t^2(1-b^2)+b^2)^2} \geq 0,$$

we have $\xi^* = \frac{SB_{in} b^2}{t^2(1-b^2)+b^2} = \zeta^*$. So it remains to show that $\mathbf{y}^* = (y_0^*, y_1^*, y_2^*)$ is feasible for all $b \in [0,1]$. Let $g(a) := a y_1^* + a^2 y_2^* - r_b(a)$. By the following observations

$$\frac{d^3 g}{da^3} = \frac{24ab^4(1-b^2)}{(a^2(1-b^2)+b^2)^4}(b^2 - a^2(1-b^2))$$
$$\frac{d^3 g}{da^3} \geq 0 \quad \text{if } a \in [0, \sqrt{\frac{b^2}{1-b^2}}]$$
$$g(0) = 0$$
$$g(t) = 0$$
$$\left.\frac{dg}{da}\right|_{a=t} = 0,$$

we have $g(a) \geq 0$ for all $b \in [0,1]$, $a \in [0, \sqrt{\frac{b^2}{1-b^2}}]$, which can be proved by exactly the same argument as in the proof of *Lemma 6*. We then consider the case $a \in [\sqrt{\frac{b^2}{1-b^2}}, 1]$. Using a similar argument based on the MVT as in *Lemma 6*, we can prove $g'\left(\sqrt{\frac{b^2}{1-b^2}}\right) \geq 0$ by first showing $g''(a) \geq 0, \forall a \in [t, \sqrt{\frac{b^2}{1-b^2}}]$. By noting that $g''(a) = 2y_2^* + 2b^4 \frac{3a^2(1-b^2)-b^2}{(a^2(1-b^2)+b^2)^3} \geq 0$ for all $a \in [\sqrt{\frac{b^2}{1-b^2}}, 1]$, we conclude that $g'(a) \geq 0$ for all $a \in [\sqrt{\frac{b^2}{1-b^2}}, 1]$. Since $g\left(\sqrt{\frac{b^2}{1-b^2}}\right) \geq 0$ has been proved in the first case, we then have $g(a) \geq 0$ for all $a \in [\sqrt{\frac{b^2}{1-b^2}}, 1]$. From the above reasoning, we have $g(a) \geq 0$ for all $a, b \in [0,1]$, and thus $\mathbf{y}^*$ is feasible and the proposition follows. ∎

From *Propositions 2* to *4*, we have the following tight upper bound:

$$\int r_b(a) dP^*(a) = \int \frac{a^2 b^2}{a^2(1-b^2) + b^2} dP^*(a)$$
$$= s(CB_{in}, SB_{in}, b),$$

where

$$s(CB_{in}, SB_{in}, b)$$
$$= \begin{cases} \frac{CB_{in}^2 b^2}{CB_{in}^2(1-b^2)+b^2} & \text{if } b \in \left[0, \sqrt{\frac{CB_{in}^2}{1+CB_{in}^2}}\right] \\ \frac{CB_{in}}{2}\sqrt{\frac{b^2}{1-b^2}} & \text{if } b \in \left[\sqrt{\frac{CB_{in}^2}{1+CB_{in}^2}}, \sqrt{\frac{t^2}{1+t^2}}\right] \\ \frac{SB_{in} b^2}{t^2(1-b^2)+b^2} & \text{if } b \in \left[\sqrt{\frac{t^2}{1+t^2}}, 1\right] \end{cases}. \quad (49)$$

Hereafter, we will show that the $b$-value-independent $dP^{**}(a)$ in (34) is an upper bounding distribution, such that $dP^{**}(a)$ may not be feasible, but the resulting $\int r_b(a) dP^{**}(a)$ is no smaller than $s(CB_{in}, SB_{in}, b)$ for all $b \in [0,1]$.

*Lemma 7:* $\int r_b(a) dP^{**}(a) \geq s(CB_{in}, SB_{in}, b)$ for all $b \in [0, \sqrt{\frac{CB_{in}^2}{1+CB_{in}^2}}]$.

*Proof:* By the monotonicity of $r_b(a)$ as a function of $a$, we have

$$\int r_b(a)dP^{**}(a)$$
$$= (1-f_{SB})\frac{t}{t+CB_{in}}r_b(CB_{in}) + f_{SB}r_b(\sqrt{SB_{in}})$$
$$+ (1-f_{SB})\frac{CB_{in}}{t+CB_{in}}r_b(t)$$
$$\geq (1-f_{SB})\frac{t}{t+CB_{in}}r_b(CB_{in}) + f_{SB}r_b(CB_{in})$$
$$+ (1-f_{SB})\frac{CB_{in}}{t+CB_{in}}r_b(CB_{in})$$
$$= r_b(CB_{in}) = s(CB_{in}, SB_{in}, b).$$

■

*Lemma 8:* $\int r_b(a)dP^{**}(a) \geq s(CB_{in}, SB_{in}, b)$ for all $b \in [\sqrt{\frac{CB_{in}^2}{1+CB_{in}^2}}, \sqrt{\frac{t^2}{1+t^2}}]$.

*Proof:* We prove this by directly applying calculus. By changing variables to $x := \sqrt{\frac{b^2}{1-b^2}}$ and using $c$ as a shortcut of $CB_{in}$ (note that $SB_{in} = tc$), proving *Lemma 8* is equivalent to showing

$$\int r_b(a)dP^{**}(a)$$
$$= (1-f_{SB})\frac{t}{t+c}\frac{c^2x^2}{c^2+x^2} + f_{SB}\frac{tcx^2}{tc+x^2}$$
$$+ (1-f_{SB})\frac{c}{t+c}\frac{t^2x^2}{t^2+x^2}$$
$$= tcx^2\left(\frac{x^4+t^2c^2+x^2\left(t^2+c^2-(1-f_{SB})(t-c)^2\right)}{(x^2+t^2)(x^2+tc)(x^2+c^2)}\right)$$
$$\geq s(CB_{in}, SB_{in}, b)$$
$$= \frac{c}{2}x, \qquad \forall x \in [c,t] \subseteq [0,1].$$

Multiplying the common denominator and changing the variable to $y := \frac{x}{\sqrt{tc}}$, the desired inequality becomes

$$t^3c^3y^6 + t^2c^2(t^2+c^2+tc)y^4 + t^2c^2(t^2+c^2+tc)y^2 + t^3c^3$$
$$- 2t\sqrt{tc}y(t^2c^2(y^4+1) + tcy^2(t^2+c^2-(1-f_{SB})(t-c)^2))$$
$$\leq 0,$$

for all $\frac{\sqrt{c}}{\sqrt{t}} < y \leq \frac{\sqrt{t}}{\sqrt{c}}$. By again changing the variable to $w := \sqrt{tc}(y+\frac{1}{y})$, we would like to prove that

$$\eta(w) - 2f_{SB}t(t-c)^2 \leq 0, \quad \forall w \in [2\sqrt{tc}, (t+c)],$$

where $\eta(w)$ is defined in (35). By noting that $\eta(t+c) - 2f_{SB}t(t-c)^2 = -2(t-c)(c(t+c)+ft(t-c)) \leq 0$ for all $f_{SB} \in [0,1]$, we would like to show that there exists no root of $\eta(w)-2f_{SB}t(t-c)^2$ in $[2\sqrt{tc}, t+c]$. If $t - \sqrt{c(2t-c)} \leq 2\sqrt{tc}$, then by definition $f_{SB} = 0$. By simple calculus, there is no root in $[2\sqrt{tc}, t+c]$. If $2\sqrt{tc} - t + \sqrt{c(2t-c)} < 0$, there is one root of $\eta(w)$ in $[2\sqrt{tc}, t+c]$. By letting $f_{SB} = \frac{\eta(w^*)}{2t(t-c)^2}$ where

$$w^* = \begin{cases} 2\sqrt{tc} & \text{if } \eta'(2\sqrt{tc}) \leq 0 \\ \frac{2t-\sqrt{4t^2-3(t-c)^2}}{3} & \text{otherwise} \end{cases},$$

we guarantee that $\eta(w) - 2f_{SB}t(t-c)^2$, the shifted version of $\eta(w)$, has no root in $[2\sqrt{tc}, t+c]$. This completes the proof. ■

*Lemma 9:* $\int r_b(a)dP^{**}(a) \geq s(CB_{in}, SB_{in}, b)$ for all $b \in (\sqrt{\frac{t^2}{1+t^2}}, 1]$.

*Proof:* In this proof, we use another index $\alpha \triangleq a^2 = 4p(1-p)$ for different BSCs and $dP(\alpha)$ now denotes the corresponding probabilistic weight for BSCs indexed by $\alpha$, which is different from the $dP(p)$ and $dP(a)$ discussed previously. We can rewrite $r_b(a)$ with respect to the new index $\alpha$ such that it becomes $r_b(\alpha) = \frac{\alpha b^2}{\alpha(1-b^2)+b^2}$, the $SB$ value when the constituent BSCs are indexed by $\alpha$ and $b$ respectively.

It can be shown that $\frac{d^2 r_b(\alpha)}{d\alpha^2} \leq 0$ and $r_b(\alpha)$ is a concave function of $\alpha$. By noting that $\int \alpha dP^{**}(\alpha) = SB_{in}$ and the weights in $dP^{**}(\alpha)$ are concentrated only on three points $\alpha = CB_{in}^2$, $SB_{in}$, and $t^2$ in an increasing order, we have $\int r_b(\alpha)dP^{**}(\alpha) \geq A_1$, where $A_1$ is the intersection of the vertical line $\alpha = SB_{in}$ and the chord connecting $(CB_{in}^2, r_b(CB_{in}^2))$ and $(t^2, r_b(t^2))$.

We also notice that $s(CB_{in}, SB_{in}, b)$ is the intersection of the vertical line $\alpha = SB_{in}$ and the chord connecting $(0, r_b(0))$ and $(t^2, r_b(t^2))$. By the concavity of $r_b(\alpha)$, we conclude $\int r_b(\alpha)dP^{**}(\alpha) \geq A_1 \geq s(CB_{in}, SB_{in}, b)$. ■

APPENDIX VI
PROOF OF *Theorem 8*

We provide a proof of a more general theorem, which includes general error correcting codes and multiuser detection as special cases and is formally stated as follows.

As in Fig. 9, consider any *deterministic/randomized* sequence mapper[12] $C : \{0,1\} \mapsto \{0,1\}^n$ and $\mathbf{W} = C(X)$. Each coordinate $W_i$ of $\mathbf{W}$ is passed through independent BI-SO channels $F_i := \{f_i(y|w)\}$ to generate the observation $Y_i$, $i \in \{1, 2, \cdots, n\}$. Let $\hat{X}(\mathbf{Y})$ be the MAP detector, and define

$$p_e(\{F_i\}) := \mathsf{P}_{X,\mathbf{Y}}(\hat{X}(\mathbf{Y}) \neq X)$$
$$\text{and } CB(\{F_i\}) := \mathsf{E}_{X,\mathbf{Y}}\left\{\sqrt{\frac{\mathsf{P}(\bar{X}|\mathbf{Y})}{\mathsf{P}(X|\mathbf{Y})}}\right\}$$

as the error probability and $CB$ value of this $X \mapsto \mathbf{Y}$ vector channel given the conditional channel distributions $\{F_i\}$. We then have the following theorem.

*Theorem 9:* For any *uniform/nonuniform* binary input distribution on $X$, we have

$$CB(\{F_i\}) \leq CB(\{F_{BSC, \tilde{p}_i}\}),$$

where for any $i$, $\tilde{p}_i$ satisfies $4\tilde{p}_i(1-\tilde{p}_i) = \int 4p(1-p)dP_i(p)$. The integrator $dP_i(p)$ is the equivalent probabilistic weight in the BSC decomposition of channel $F_i$ as described in Section II-A.2.

*Theorem 8* is a special case of *Theorem 9* obtained by letting $X \mapsto C(X)$ be the binary-input/vector-output support tree channel.

---
[12]$X$ and $C(X)$ can be regarded as a binary-input vector-output channel. Or $C(X)$ is the subspace of codewords corresponding to information bit $X$.

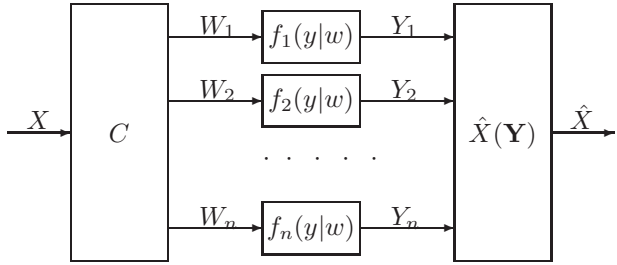

Fig. 9. General deterministic/randomized bit to sequence mapper with independent symmetric channels.

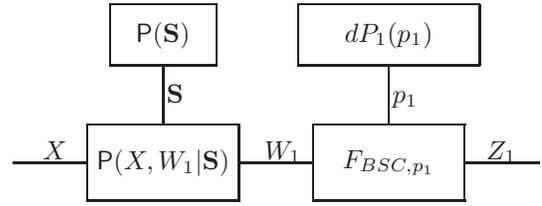

Fig. 10. The factor graph of the five random variables: $X, Z_1, W_1, \mathbf{S}$, and $p_1$.

*Note 1:* In the setting of *Theorem 9*, we only require all constituent channels to be of BI-SO type. The bit-to-sequence mapper $X \mapsto C(X)$ does not need to be symmetric, which is different from the case of LDPC codes.

*Note 2:* The definition of $SB$ in (2) is valid for general BI-NSO channels with arbitrary input distributions. However, with a non-uniform input distribution, $SB(F_{BSC,p}) \neq 4p(1-p)$. This is the reason why in *Theorem 9*, we deliberately use $4\tilde{p}_i(1-\tilde{p}_i) = \int 4p(1-p) dP_i(p)$ instead of $SB(F_{BSC,\tilde{p}_i}) = SB(F_i)$.

*Proof of Theorem 9:* By rewriting each BI-SO channel $F_i$ as the probabilistic combination of BSCs with weights $dP_i(p_i)$, each observation $y_i$ can be viewed as a pair $(z_i, p_i) \in \{0,1\} \times [0, 1/2]$, where $z_i$ is the binary output of $F_{BSC,p_i}$ and $p_i$ is the side information specifying the crossover probability of the corresponding BSC. Taking the marginal approach, we will focus on $y_1 = (z_1, p_1)$ and treat all $y_2, y_3, y_4, \cdots, y_n$ as the side information $\mathbf{s}$. The conditional probability $\mathsf{P}(\cdot | \mathbf{S} = \mathbf{s}, p_1)$ can then be factored as

$$\mathsf{P}(X, W_1, Z_1 | \mathbf{S} = \mathbf{s}, p_1)$$
$$= (p_1 + (1-2p_1)\delta(Z_1 - W_1)) \mathsf{P}(X, W_1 | \mathbf{S} = \mathbf{s}, p_1)$$
$$= (p_1 + (1-2p_1)\delta(Z_1 - W_1)) \mathsf{P}(X, W_1 | \mathbf{S} = \mathbf{s}), \quad (50)$$

where $\delta(x) = 1$ iff $x = 0$. To write $\mathsf{P}(X, W_1 | \mathbf{S}, p_1) = \mathsf{P}(X, W_1 | \mathbf{S})$, we use the fact that knowing what type of BSCs we are facing (namely, knowing $p$) provides no information[13] about the input distribution ($W_1$). This fact also implies that $dP_1(p)$ does not depend on the distribution of $\mathbf{S}$ either. As a result, we have

$$\mathsf{P}(\mathbf{S}, p_1) = \mathsf{P}(\mathbf{S})\mathsf{P}(p_1) = \mathsf{P}(\mathbf{S}) dP_1(p_1). \quad (51)$$

By (50) and (51), the corresponding factor graph is drawn in Fig. 10. We can rewrite the conditional distribution $\mathsf{P}(X, W_1 | \mathbf{S})$ in the matrix form:

$$\begin{pmatrix} \mathsf{P}(W_1=0, X=0 | \mathbf{S}=\mathbf{s}) & \mathsf{P}(W_1=0, X=1 | \mathbf{S}=\mathbf{s}) \\ \mathsf{P}(W_1=1, X=0 | \mathbf{S}=\mathbf{s}) & \mathsf{P}(W_1=1, X=1 | \mathbf{S}=\mathbf{s}) \end{pmatrix}$$
$$= \begin{pmatrix} a & b \\ c & d \end{pmatrix},$$

where $a, b, c,$ and $d$ are functions of $\mathbf{s}$ satisfying $a, b, c, d \geq 0$ and $a + b + c + d = 1$. It is worth repeating that $a, b,$ and $d$ do not depend on $p_1$. The conditional input-output distribution $\mathsf{P}(X, Z_1 | \mathbf{S}, p_1)$ then becomes

$$\begin{pmatrix} \mathsf{P}(Z_1=0, X=0 | \mathbf{S}=\mathbf{s}, p_1) & \mathsf{P}(Z_1=0, X=1 | \mathbf{S}=\mathbf{s}, p_1) \\ \mathsf{P}(Z_1=1, X=0 | \mathbf{S}=\mathbf{s}, p_1) & \mathsf{P}(Z_1=1, X=1 | \mathbf{S}=\mathbf{s}, p_1) \end{pmatrix}$$
$$= \begin{pmatrix} a(1-p_1) + cp_1 & b(1-p_1) + dp_1 \\ ap_1 + c(1-p_1) & bp_1 + d(1-p_1) \end{pmatrix}.$$

---

[13]$dP(p)$ only depends on the channel distribution $f(z|w)$, not on the *a priori* distribution of $W$. This is a special property of the BSC decomposition mentioned in Section II-A.2. For BI-NSO channels, though the corresponding BNSC decomposition can be found as in Section V-A, the probabilistic weight $dP(p_{0 \to 1}, p_{1 \to 0})$ depends on the distribution of $W$.

The value of $CB$ for the $X \mapsto \mathbf{Y}$ channel (or equivalently $X \mapsto (Z_1, p_1, \mathbf{S})$) becomes $CB = \mathsf{E}\left\{\sqrt{\frac{\mathsf{P}(\bar{X}|Z_1, p_1, \mathbf{S})}{\mathsf{P}(X|Z_1, p_1, \mathbf{S})}}\right\}$. Taking the expectation step by step, we have

$$CB_{X, Z_1 | p_1, \mathbf{S}}$$
$$:= \mathsf{E}_{X, Z_1 | p_1, \mathbf{S}} \left\{\sqrt{\frac{\mathsf{P}(\bar{X}|Z_1, p_1, \mathbf{S})}{\mathsf{P}(X|Z_1, p_1, \mathbf{S})}}\right\}$$
$$= 2\sqrt{(a(1-p_1) + cp_1)(b(1-p_1) + dp_1)}$$
$$+ 2\sqrt{(ap_1 + c(1-p_1))(bp_1 + d(1-p_1))}.$$

By *Proposition 5* (stated at the end of this proof), $CB_{X, Z_1 | p_1, \mathbf{S}}$ is a concave function of $\beta := 4p_1(1-p_1)$ for all valid $a, b, c,$ and $d$. By Jensen's inequality, for any channel $F_C$,

$$CB_{X, Z_1, p_1 | \mathbf{S}} := \mathsf{E}_{X, Z_1, p_1 | \mathbf{S}} \left\{\sqrt{\frac{\mathsf{P}(\bar{X}|Z_1, p_1, \mathbf{S})}{\mathsf{P}(X|Z_1, p_1, \mathbf{S})}}\right\}$$
$$= \int_{p=0}^{1/2} CB_{X, Z_1 | p_1, \mathbf{S}} dP_1(p_1)$$
$$\leq CB_{X, Z_1 | \tilde{p}_1, \mathbf{S}}, \quad (52)$$

where $\tilde{p}_1$ is the crossover probability such that $4\tilde{p}_1(1-\tilde{p}_1) = \int 4p_1(1-p_1) dP_1(p_1)$. By (52) and noting that $F_{BSC,\tilde{p}_1}$ is the universal maximizing distribution for any realization of $\mathbf{S}$, we obtain that

$$CB(\{F_1, F_2, \cdots, F_n\}) = \int CB_{X, Z_1, p_1 | \mathbf{S}} d\mathsf{P}(\mathbf{S})$$
$$\leq \int CB_{X, Z_1 | \tilde{p}_1, \mathbf{S}} d\mathsf{P}(\mathbf{S})$$
$$= CB(\{F_{BSC,\tilde{p}_1}, F_2, F_3, \cdots, F_n\}).$$

By repeatedly applying this $CB$-increasing channel replacement until all constituent channels $F_i$ are replaced by $F_{BSC,\tilde{p}_i}$, the proof of *Theorem 9* is complete. ∎

*Proposition 5:* For any constants $a, b, c, d \geq 0$ and $p \in [0, 1/2]$, we have

$$\sqrt{(a(1-p) + cp)(b(1-p) + dp)}$$
$$+ \sqrt{(ap + c(1-p))(bp + d(1-p))}$$

is a concave function of $\beta := 4p(1-p)$.

*Proof:* This proof involves several changes of variables. It is worth noting that this proposition is a pure algebraic statement and the notations involved herein are irrelevant to those of the LDPC code problem.

We first let $X = 1 - 2p$, $A = \frac{a+c}{|a-c|}$, and $B = \frac{b+d}{|b-d|}$. Then the problem becomes to prove that both

$$f(X) = \sqrt{A+X}\sqrt{B+X} + \sqrt{A-X}\sqrt{B-X}$$

and

$$g(X) = \sqrt{A-X}\sqrt{B+X} + \sqrt{A+X}\sqrt{B-X}$$

are concave functions of $\beta = 1 - X^2$, for all $A, B \in [1, \infty]$ and $X \in [0, 1]$. We focus on the concavity of $f(X)$ first. Using the chain rule,

$$\frac{df(X)}{d\beta} = \frac{df(X)}{dX}\frac{dX}{d\beta} = \frac{1}{2} \cdot \frac{1}{2X}$$
$$\left(-\sqrt{\frac{A+X}{B+X}} - \sqrt{\frac{B+X}{A+X}} + \sqrt{\frac{A-X}{B-X}} + \sqrt{\frac{B-X}{A-X}}\right).$$

Since $\frac{d^2 f(X)}{d\beta^2} = -\frac{1}{2X}\frac{d}{dX}\frac{df(X)}{d\beta}$, showing the concavity of $f(X)$ as a function of $\beta$ is equivalent to showing:

$$\frac{d}{dX}\frac{df(X)}{d\beta}$$
$$= \frac{1}{4X^2}\left(\frac{A+B+X+\frac{X}{2}\left(\frac{A+X}{B+X}+\frac{B+X}{A+X}\right)}{\sqrt{(A+X)(B+X)}}\right.$$
$$\left. + \frac{-A-B+X+\frac{X}{2}\left(\frac{A-X}{B-X}+\frac{B-X}{A-X}\right)}{\sqrt{(A-X)(B-X)}}\right)$$
$$\triangleq \frac{1}{4X^2}f_2(X) \geq 0.$$

To show $f_2(X) \geq 0$, we first note that $f_2(0) = 0$. Its first derivative is

$$\frac{df_2(X)}{dX} = \frac{3(A-B)^2 X}{4}$$
$$\left(\frac{(A-X)+(B-X)}{\sqrt{(A-X)(B-X)}^5} - \frac{(A+X)+(B+X)}{\sqrt{(A+X)(B+X)}^5}\right).$$

By *Lemma 10* (stated at the end of this proof), we have $\frac{df_2(X)}{dX} \geq 0$. Thus $f_2(X) \geq 0$, which implies that $f(X)$ is concave as a function of $\beta$.

For $g(X)$, we have

$$\frac{dg(X)}{d\beta} = \frac{1}{2} \cdot \frac{1}{2X}$$
$$\left(\sqrt{\frac{A+X}{B-X}} - \sqrt{\frac{B-X}{A+X}} - \sqrt{\frac{A-X}{B+X}} + \sqrt{\frac{B+X}{A-X}}\right).$$

Since $\frac{d^2 g(X)}{d\beta^2} = -\frac{1}{2X}\frac{d}{dX}\frac{dg(X)}{d\beta}$, showing the concavity of $g(X)$ is equivalent to showing

$$\frac{d}{dX}\frac{dg(X)}{d\beta}$$
$$= \frac{1}{4X^2}\left(\frac{-A+B-X+\frac{X}{2}\left(\frac{A+X}{B-X}+\frac{B-X}{A+X}\right)}{\sqrt{(A+X)(B-X)}}\right.$$
$$\left. + \frac{A-B-X+\frac{X}{2}\left(\frac{A-X}{B+X}+\frac{B+X}{A-X}\right)}{\sqrt{(A-X)(B+X)}}\right)$$
$$\triangleq \frac{1}{4X^2}g_2(X) \geq 0.$$

To show $g_2(X) \geq 0$, we first note that $g_2(0) = 0$. Its first derivative is

$$\frac{dg_2(X)}{dX} = \frac{3(A+B)^2 X}{4}$$
$$\left(\frac{(A+X)-(B-X)}{\sqrt{(A+X)(B-X)}^5} + \frac{(B+X)-(A-X)}{\sqrt{(A-X)(B+X)}^5}\right).$$

By *Lemma 10*, we have $\frac{dg_2(X)}{dX} \geq 0$, which implies that $g(X)$ is concave as a function of $\beta$. This completes the proof of *Proposition 5*. ∎

*Lemma 10:* For all $a, b, c \geq 0$, we have

$$\frac{a+b}{(ab)^{5/2}} \geq \frac{(a+c)+(b+c)}{((a+c)(b+c))^{5/2}}$$
$$\frac{(a+c)-b}{((a+c)b)^{5/2}} + \frac{(b+c)-a}{((b+c)a)^{5/2}} \geq 0.$$

*Proof:* By noting that $\frac{a+b}{ab} = \frac{1}{a} + \frac{1}{b} \geq \frac{1}{a+c} + \frac{1}{b+c} = \frac{(a+c)+(b+c)}{(a+c)(b+c)}$ and $0 \leq ab \leq (a+c)(b+c)$, we prove the first inequality.

For the second inequality, without loss of generality, we assume $a \leq b$. We then observe that

$$(a+c)b \geq (b+c)a \geq 0$$
$$\frac{(b+c)-a}{(b+c)a} \geq 0$$
$$\frac{(a+c)-b}{(a+c)b} + \frac{(b+c)-a}{(b+c)a} = \frac{1}{b} - \frac{1}{a+c} + \frac{1}{a} - \frac{1}{b+c} \geq 0. \tag{53}$$

Considering (53), after multiplying the non-negative second term $\frac{(b+c)-a}{(b+c)a}$ by a larger factor $\frac{1}{\sqrt{(b+c)a}^3}$ and the *possibly-negative* first term $\frac{(a+c)-b}{(a+c)b}$ by a smaller factor $\frac{1}{\sqrt{(a+c)b}^3}$, the new weighted sum is no less than zero, namely,

$$\frac{(a+c)-b}{((a+c)b)^{5/2}} + \frac{(b+c)-a}{((b+c)a)^{5/2}} \geq 0.$$

This completes the proof. ∎

PLACE PHOTO HERE

**Chih-Chun Wang** (M'06) received the B.E. degree in E.E. from National Taiwan University, Taipei, Taiwan in 1999, the M.S. degree in E.E., the Ph.D. degree in E.E. from Princeton University in 2002 and 2005, respectively. He worked in COMTREND Corporation, Taipei, Taiwan, from 1999-2000, and spent the summer of 2004 with Flarion Technologies. In 2005, he was a postdoctoral researcher with Princeton University working on channel coding and graphical algorithms. Since 2006, he has joined Purdue University, where he is an assistant professor in the School of Electrical and Computer Engineering. His research interests are in optimal control, information theory, coding theory, iterative inference algorithms, graphical algorithms, and network coding.

PLACE PHOTO HERE

**Sanjeev R. Kulkarni** (M'91, SM'96, F'04) received the B.S. in Mathematics, B.S. in E.E., M.S. in Mathematics from Clarkson University in 1983, 1984, and 1985, respectively, the M.S. degree in E.E. from Stanford University in 1985, and the Ph.D. in E.E. from M.I.T. in 1991.

From 1985 to 1991 he was a Member of the Technical Staff at M.I.T. Lincoln Laboratory working on the modelling and processing of laser radar measurements. In the spring of 1986, he was a part-time faculty at the University of Massachusetts, Boston. Since 1991, he has been with Princeton University where he is currently Professor of Electrical Engineering. He spent January 1996 as a research fellow at the Australian National University, 1998 with Susquehanna International Group, and summer 2001 with Flarion Technologies.

Prof. Kulkarni received an ARO Young Investigator Award in 1992, an NSF Young Investigator Award in 1994, and several teaching awards at Princeton University. He has served as an Associate Editor for the IEEE Transactions on Information Theory. Prof. Kulkarni's research interests include statistical pattern recognition, nonparametric estimation, learning and adaptive systems, information theory, wireless networks, and image/video processing.

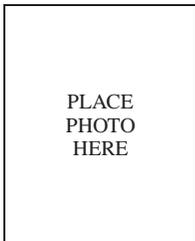

**H. Vincent Poor** (S'72, M'77, SM'82, F'87) received the Ph.D. degree in EECS from Princeton University in 1977. From 1977 until 1990, he was on the faculty of the University of Illinois at Urbana-Champaign. Since 1990 he has been on the faculty at Princeton, where he is the Dean of Engineering and Applied Science, and the Michael Henry Strater University Professor of Electrical Engineering. Dr. Poor's research interests are in the areas of stochastic analysis, statistical signal processing and their applications in wireless networks and related fields. Among his publications in these areas is the recent book MIMO Wireless Communications (Cambridge University Press, 2007).

Dr. Poor is a member of the National Academy of Engineering and is a Fellow of the American Academy of Arts and Sciences. He is also a Fellow of the Institute of Mathematical Statistics, the Optical Society of America, and other organizations. In 1990, he served as President of the IEEE Information Theory Society, and he is currently serving as the Editor-in-Chief of these Transactions. Recent recognition of his work includes a Guggenheim Fellowship and the IEEE Education Medal.